# Performance-Based Optimization of 2D Reinforced Concrete Moment Frames through Pushover Analysis and ABC Optimization Algorithm

*Saba Faghirnejad*[a], Mohammadreza *Ghasemi*[b]

[a] Lecturer, Department of Civil Engineering, The University of Lorestan, Khorramabad, Lorestan, Iran

[b] Professor, Department of Civil Engineering, University of Sistan and Baluchestan, Zahedan, Iran

1. Introduction
   1.1. Background and Importance of Seismic Design for Reinforced Concrete Moment Frames
   1.2. Introduction to Performance-Based Design and its Significance in Seismic Engineering
   1.3. Application of Optimization in Seismic Engineering
   1.4. Research Objectives and Scope of the Study
   1.5. Research Gap
   1.6. Outline of the Paper

2. Research Method
   2.1. Performance-Based Design and Pushover Analysis
   2.2. Overview of the ABC Optimization Algorithm for Structural Design
   2.3. Integration of Performance-Based Design, Pushover Analysis, and ABC Optimization
   2.4. Design Parameters and Constraints for the Study
   2.5. Overall Methodology

3. Case Studies
   3.1. Description of the 2D Reinforced Concrete Moment Frames in the Study
   3.2. Application of Performance-Based Design, Pushover Analysis, and ABC Optimization to the Case Studies
   3.3. Performance Indices for Seismic Assessment

4. Results and Discussion
   4.1. Comparative Analysis of Different Seismic Design Configurations
   4.2. Identification of the Optimal Design using ABC Optimization
   4.3. Performance Assessment of the Optimized 2D Moment Frames
   4.4. Sensitivity Analysis of Design Parameters

5. Conclusion
   5.1. Summary of Findings from the Study
   5.2. Significance of Performance-Based Design and ABC Optimization in Enhancing Seismic Performance
   5.3. Practical Applications and Recommendations for Future Research

**Abstract**

Conducting nonlinear pushover analysis typically demands intricate and resource-intensive computational attempts, and involves a process that is highly iterative and necessary for satisfying design-defined and also requirements of codes in performance-based design. A computer-based technique is presented for reinforced concrete (RC) buildings in this study, incorporating optimization numerical approaches, techniques of optimality criteria and pushover analysis to seismic design automatically the pushover drift performance.

The optimal design based on the performance of concrete beams, columns and shear walls in concrete moment frames is presented using the artificial bee colony optimization algorithm. The design is applied to three frames such as a 4-story, an 8-story and a 12-story. These structures are designed to minimize the overall weight while satisfying the levels of performance include Life Safety (L-S), Collapse Prevention (C-P), and Immediate Occupancy (I-O). To achieve this goal, three main steps are performed. In the first step, optimization codes are implemented in MATLAB software, and the OpenSees software is used for nonlinear static analysis of the structure. By solving the optimization problem, several top designs are obtained for each frame and shear wall. Pushover analysis is performed considering the constraints of relative displacement and plastic hinge rotation based on the nonlinear provisions of FEMA356 code to achieve each levels of performance. Following this, convergence, pushover, and drift history curves are plotted for each frame, and selecting the best design for each frame ultimately occurs. The results demonstrate the algorithm's performance is desirable for the structure to achieve selecting the best design and lower weight.

## 1. Introduction

Earthquakes, as significant natural disasters, pose threats to both human lives and built infrastructure [1]. The damages caused by seismic forces on engineering structures are undeniable. However, the detrimental impact can be mitigated through meticulous attention to seismic design provisions. Designing structures to withstand seismic loads often leads to constructions with elevated costs. Therefore, optimizing the cost of construction while simultaneously satisfying design criteria becomes a rational endeavor. Structural optimization offers a pathway to achieve economical designs. Within the realm of seismic design, the objective function plays a central role in shaping and guiding the optimization process. This function frequently revolves around two key aspects: the weight or cost of structures [2]. The inherent connection between weight and cost underscores the importance of minimizing structural weight, as a decrease in weight is directly associated with an overall reduction in cost. Because of several factors like the materials' heterogeneous nature and the presence of numerous sizes and configurations of reinforcement members, achieving the optimal design of the reinforced elements in structures, particularly when compared to steel structures, is complex [3]. Structural elements known as shear walls, or structural walls, are integral components that primarily bear lateral loads arising from factors like wind and seismic activities. These walls frequently function as lateral bracing elements for the entire structure. They bear the weight imposed by components connected to the wall and are responsible for resisting lateral shear forces and moments around the wall's principal axis [4]. In areas characterized by medium to high seismic risk levels, it is essential to implement specific reinforcement measures as highlighted by design codes. These measures are necessary to guarantee that concrete structures exhibit satisfactory performance in the face of seismic hazards [5].

The significant influence of earthquakes in seismic-prone areas highlights the critical importance of preserving the safety and integrity of structures. From the 1990s onwards, catalyzed by seismic events like the 1989 Loma Prieta and Northridge earthquakes, there has been a growing inclination, particularly in the United States, towards integrating performance-based seismic engineering (PBSE) into building design methodologies. This evolution is marked by the alignment of seismic design regulations with the principles of performance-based design (PBD), signifying a contemporary trajectory within the realm of structural engineering. This shift towards PBD accentuates the significance of employing nonlinear analysis methodologies to comprehend damage patterns and magnitudes in structures. Particularly crucial in assessing inelastic behaviors and failure modes during intense seismic events, performance-based design aims to ensure structures meet predefined performance benchmarks across specific hazard scenarios. Unlike traditional design approaches reliant on predetermined rules, PBD seeks to achieve specific performance levels in line with the intended functionality, safety, and serviceability of structures. This design philosophy empowers structural engineers to define the desired performance standard for structures and their corresponding hazard thresholds, steering the design process towards predetermined seismic outcomes. A central tenet of PBD revolves around optimizing material utilization to achieve cost-effective solutions. Among these methodologies, pushover analysis emerges as a simplified yet robust procedure that considers nonlinearity. This technique involves the incremental application of a predefined sequence of seismic loads to structural systems until a plastic collapse mechanism becomes evident. Advancements in mathematical and evolutionary algorithms have significantly propelled the pursuit of performance-based optimization design (PBOD) in the field of structural engineering [6], [7], [8], [9], [10].

The nonlinear static analysis process comprises three fundamental elements: performance, demand, and capacity. An estimate of the structure's behavior beyond the elastic limit when subjected to seismic loads is provided by the capacity spectrum, which is generated via pushover analysis primarily utilizing the structure's first-mode response. On the other hand, adjusting the conventional elastic design spectrum with a 5% damping ratio results in the derivation of the demand spectrum curve. The "point of performance," a pivotal threshold where structural responses must meet defined acceptability criteria, is established when the pushover demand and capacity spectrum curves intersect. These responses are assessed against predefined acceptability limits at both the global system and local element levels, taking into account aspects like inter-story drift and lateral load stability. If a structure's responses fall short of the targeted performance, iterative design processes are required to achieve the desired level of performance, even with modern engineering software. The importance of inter-story drift performance in multistory buildings, serving as a metric for structural and non-structural damage under diverse earthquake motions, is widely recognized. The PBD accords inter-story drift as a principal criterion, assessing system performance through drift values along the building's height under various seismic conditions. Ensuring uniform ductility across all stories through inter-story drift control is crucial to averting catastrophic collapses. However, economically designing building elements to accommodate diverse levels of elastic and inelastic drift performance remains a challenging task [6], [7] ,[8], [9], [10].

Structural optimization holds a pivotal role in seismic engineering, contributing to the creation of cost-effective and resilient designs capable of withstanding seismic forces. Optimization techniques aim to identify optimal configurations of structural elements, materials, and dimensions to meet predetermined performance objectives These objectives may encompass minimizing construction costs, maximizing structural strength, or reducing damage under seismic loads. Whether based on mathematical algorithms or heuristic approaches, optimization methods provide engineers with powerful tools to explore the design space and find optimal solutions that strike a balance between conflicting design requirements [9], [10], [11].

While various researchers have delved into optimizing the design there is a noticeable gap in the literature concerning the shear walls' seismic optimization that incorporates the most up-to-date seismic design standards. For instance, in their work, Saka [12] introduced an algorithm designed to optimize multi-floor reinforced concrete structures incorporating shear walls which enhances structural design by taking into account various factors, including bending moment, displacement, minimum size constraints and ultimate axial load. Ganzerli [13] et al , introduce a cost-effective seismic design approach, combining structural optimization with performance-based criteria, influenced by retrofitting guidelines, to ensure actual structural performance during earthquakes and achieve quantifiable reliability levels in the design. Cheng and Pantelides [14] conducted research on optimal actuator placement for seismic structural control, aiming to minimize structural response through three methods for optimal location selection. Fragiadakis and Papadrakakis [15] explored Performance-Based Optimization and Design (PBOD) within an automated structural design framework, providing deterministic or probabilistic solutions with diverse objectives and limit states. Grierson & Moharrami [16] developed an efficient computer-based method for cost-effective, reinforced concrete building framework design. Fadaee and Grierson [17] introduced a computer-based optimization technique for 3D reinforced concrete structures employing the OC method. Their main objective was cost minimization while ensuring alignment with ACI code [18] regulations, which involved sensitivity analysis and an illustrative example. Three-dimensional reinforced concrete frames were optimized by Balling and Yao [19], who applied a simplified method to different frame configurations under various load combinations and introduced it as the most efficient approach. Rajeev and Krishnamoorthy [20] introduced a methodology based on genetic algorithms for optimizing the reinforced concrete plane frame design and addressing practical construction issues to generate rational optimal solutions. Sberna et al [21] present a framework based on genetic algorithms for optimizing seismic retrofitting in reinforced concrete frames, with a focus on cost reduction and the indirect consideration of expected annual loss reduction, using static pushover analyses and a 3D fiber-section model in a real-world case study. Alavi et al [22] present a novel seismic performance optimization approach in their research. They simultaneously minimize structural mass and controlled system energy, challenging conventional sequential methods. The study, employing a variable neighborhood search (VNS) metaheuristic method, demonstrates the superior dynamic performance of the proposed simultaneous optimization over sequentially optimized cases in shear-type frames subjected to strong ground motions. Alemu et al. [23] introduced a priority concept, formulated priority criteria (PC), and proposed an enhanced Particle Swarm Optimization variant called Priority Criteria PSO (PCPSO) to enhance the optimization process for structural design problems, particularly focusing on 2D reinforced concrete frames. Mirrashid et al [24] introduce an innovative computational intelligence model for estimating the flexural capacity of reinforced concrete-filled composite plate shear walls. Through validation against laboratory data and the Transit Search optimization algorithm, the proposed approach demonstrates superior efficacy compared to existing standards, offering a reliable framework for predicting the flexural behavior of composite shear walls. [25]Ghasemi et al. [26] explored the seismic performance of self-centering monolithic rocking walls, finding that an axial stress ratio between 0.10 and 0.15 is essential for meeting desired performance levels and suggesting specific limits for damage ratios and response modification factors. Kaveh and Rezazadeh Ardebili [27] utilize an enhanced algorithm for optimizing plasma generation to optimize 3D multi-story reinforced concrete structure design, with a focus on minimizing framework, steel, and concrete costs while satisfying ACI 318 and ASCE 7 requirements, demonstrating the algorithm's effective performance. Banerjee et al. [4] introduced a framework for optimizing shear wall placement in a 'C' shaped reinforced concrete structure to reduce plan irregularity-induced torsional effects. Their analysis of different shear wall locations in a 15-story 'C' shaped building, considering various structural parameters, offers valuable insights for the construction industry in India and worldwide. Kashani and Camp [28] evaluated four multi-objective optimization methods for designing reinforced concrete cantilever retaining walls, considering factors such as geotechnical stability, structural strength, cost, and weight. They found that NSGA-II excelled in coverage, SPEA2 and MOPSO performed well in diversity, and NSGA-II and MVO ranked higher in hypervolume based on different design variations.

This research aims to present an efficient approach to optimizing 2D reinforced concrete with moment-resisting frames. The optimization problem is formulated, employing the algorithm known as the Artificial Bee Colony (ABC) as a meta-heuristic optimizer under seismic loads within the framework of performance-based design (PBD). This method involves constructing columns, beams, and shear wall databases according to ACI criteria, followed by presenting formulations for the seismic design optimization of RC moment frames. The procedure incorporates both ordinary and effective seismic design constraints, including criteria for columns, beams, shear walls, and other seismic provisions. The paper incorporates key provisions from FEMA and ACI into the procedure. The objective function of the structure is determined by its weight.

While previous studies have extensively investigated various methods for optimizing the seismic performance of reinforced concrete structures, including pushover analysis, genetic algorithms, and performance-based design, there is a notable research gap in the context of optimizing two-dimensional reinforced concrete moment frames with shear walls by ABC algorithm. None of the mentioned studies have specifically focused on incorporating the ABC algorithm to optimize the seismic design of these specific structural configurations.

The novelty in this study stems from its amalgamation of the ABC algorithm with the optimization of shear walls and moment-frame elements in 2D reinforced concrete structures, subject to seismic loads. In this paper, the primary emphasis of the objective function pertains to the weight of structures, marking a departure from the prevalent focus on cost within the domain of numerous other investigations. Furthermore, while prior research endeavors have harnessed various optimization methodologies, including genetic algorithms, chaotic optimization, and charged system search, the application of the ABC algorithm to optimize these specific structural configurations constitutes a pioneering approach. The principal objective of this research is to redress this conspicuous research void by proffering an efficacious methodology for the optimization of these structural systems. This approach encompasses the integration of both ordinary and effective seismic design constraints, drawing insights from pivotal provisions outlined in FEMA and ACI codes. The utilization of the ABC algorithm and its underlying thrust towards minimizing structural weight injects a facet of innovation into the optimization process. Consequently, this innovation allows exploring objective functions that may be non-continuous or non-differentiable, thus expanding the horizons for reinforced concrete structures subjected to seismic loads that are evaluated within the area of PBD.

The format of the paper is delineated in the following sections: Section 2 presents the research methodology, outlining the integration of the ABC algorithm and nonlinear analysis techniques to optimize 2D reinforced concrete moment frames under seismic conditions. Section 3 provides case studies that illustrate the application of the proposed methodology. In Section 4, the results of the case studies are presented and discussed, highlighting the effectiveness of the introduced approach. Lastly, Section 5 concludes by summarizing the key findings, emphasizing the research's significance, and suggesting potential avenues for future exploration.

## 2. Research Method

### 2.1. Performance-Based Design and Pushover Analysis

The primary goal of PBD is to prevent a structure from surpassing a predetermined level of damage under various earthquakes and a specific level of confidence. Different techniques exist for designing reinforced concrete frames. In this particular method, the initial design of the structure takes into account gravity loads, and its seismic performance is subsequently assessed [29]. If the intended level of performance is not met, modifications to the dimensions of the structural members and the amount of reinforcement should be made until the desired level is reached [30]. To evaluate the structure in PBD, a design objective must be chosen, comprising one or more levels of performance and earthquake hazard. In this paper, three levels of performance, namely collapse prevention (C-P), immediate occupancy (I-O) and life safety (L-S), are considered. The corresponding hazard levels are defined with a probability of exceedance less than 2% in 50 years as the maximum considered earthquake (MCE), and with a probability of exceedance less than 10% in 50 years as the design earthquake (DE), respectively. Performance-based design can utilize various analysis techniques. The procedure in nonlinear dynamic analysis is especially advantageous, but the Procedure in nonlinear static analysis is employed to minimize computational expenses. The displacement target for each level of performance is determined using the following equations, based on FEMA-356 [8], [41], [42]:

$$\delta_t = C_0 C_1 C_2 C_3 S_a \frac{T_e^2}{4\pi^2} g \qquad (1)$$

$C_0$ = The factor of modification correlates the multi-degree-of-freedom system's roof displacement with the equivalent single-degree-of-freedom system's spectral displacement.

$C_1$ = The proportion of expected maximum displacements of inelastic to elastic.

$C_2$ = The strength deterioration, stiffness and hysteretic shape impact, on the maximum response of displacement.

$C_3$ = A modification factor determined to consider the impact of dynamic P-Δ effects, leading to an increase in displacement.

$T_e$ = The basic effective period, determined through the application of Equation (2).

$S_a$ = The acceleration of the response spectrum at the fundamental effective period.

$$T_e = T_i\sqrt{\frac{K_i}{K_e}} \tag{2}$$

$K_i$ and $K_e$ = The stiffness of the elastic and effective lateral of the building, respectively.

Presenting a highly effective approach for optimization based on the performance design in concrete moment resistance frames with shear walls is the primary objective of this study.

### 2.2. Overview of the ABC Optimization Algorithm for Structural Design

In this study, minimizing the weight of the concrete frame and satisfying the seismic performance levels based on FEMA356 is the primary aim of optimization and is formulated as follows [31], [32], [41]:

$$\begin{cases} Find & X; & X_j \in R^d & j = 1, \dots, n \\ to & Minimize & \Phi(X) & \\ Subject & to & g_i(X,t); & i = 1,..,n \end{cases} \tag{3}$$

n = Design variables number.

m = Constraints number.

X = Input variables.

Φ(X) = Objective function.

**The ABC Model**

The ABC algorithm is a swarm-based metaheuristic optimization technique inspired by swarm intelligence principles. It is inspired by the behavior of foraging honeybee colonies and is specifically developed for the optimization of numerical problems. This algorithm simulates and classifies bee behavior into three categories: scout bees, employed bees (forager bees) and onlooker bees (observer bees). A potential solution within the optimization problem is represented by every food source [33].

The ABC algorithm, a nature-inspired optimization method, harnesses the collective behavior of artificial bees to replicate the foraging actions observed in real honeybee colonies. Falling under the swarm intelligence category, it aims to find optimal solutions for complex problems. What makes the ABC algorithm particularly appealing is its ability to operate without requiring specialized knowledge about the specific problem at hand. Instead, a population of artificial bees is relied upon to collectively undergo exploration of the space of solution, adapt, and eventually converge towards a global optimum [34].

**The Comparison Engineering Design and ABC Algorithm**

The ABC Algorithm characterizes the analogy with engineering design by emulating the foraging behavior observed in bee colonies. Just as engineers navigate complex design spaces to identify optimal solutions, the ABC algorithm traverses an extensive search space to pinpoint the most advantageous solution [35]. Within the context of this analogy, engineering design challenges can be analogously likened to the spatial distribution of "flowers" within a metaphorical "garden" of possible solutions. The primary objective of the ABC Algorithm is to discern the most efficient design, much as bees seek to locate the most abundant food source. In the realm of engineering design, the evaluation of solution quality frequently hinges on multi-faceted criteria encompassing performance, cost-efficiency, and adherence to constraints. This comparison underscores the applicability of nature-inspired algorithms, exemplified by the ABC Algorithm, in addressing intricate optimization dilemmas encountered in the realm of engineering [36].

**Control Parameters of ABC Algorithm**

The ABC algorithm's control parameters are pivotal for shaping its behavior and determining the quest for optimal solutions within a design space. These parameters have a significant impact on the algorithm's performance and can be summarized as follows [33], [37]:

1- *Bees Number in a Colony ($N_P$):* Bees are controlled to explore solutions and adapt to the complexity of the problem.
2- *The limit for improving a solution ($I_L$):* Guides employed bees in exploring solution depth and seeking alternative solutions.
3- *Maximum Number of Iterations ($I_{Max}$):* Imposes a time limit on algorithm convergence.
4- *Variable Changing Percentage (VCP):* Considers multiple variables in solution exploration.
5- *Number of Independent Runs (r):* Ensures reliability through multiple independent trials and outcome assessment [33], [37].

**Steps of ABC Algorithm**

The ABC algorithm, much like other swarm intelligence-based approaches, operates through a well-defined iterative process, which is explained in the following and can be outlined by considering its key parameters and steps [33], [35], [37].

1) Initialization: The number of sources of food equal to randomly selected solution vectors from a population. $(X_1, \dots, X_{NS})$, is established. Each solution vector $X_i$ is defined as a set of variables $X_i = \{x_{i1}, x_{i2}, \dots, x_{iD}\}$ within predefined bounds.

$$x_{ij} = x_{min\,j} + rand\,[0,1].\,(x_{max\,j} - x_{min\,j}) \quad (4)$$

For $i = 1, 2, \dots, N_s$ and $j = 1, 2, \dots, D$

"$x_{max\,j}$" And "$x_{min\,j}$" correspondingly indicate the highest and lowest limits for dimension j.

2) Employed Bees: They explore neighborhoods to find new food sources by updating variables with Equation (5). It includes selecting random indexes (k and j) to create hybrid solutions, aided by $\phi_{ij}$ between [-1, 1]. Boundary limits are adjusted with repetitions based on variable change percentage.

$$v_{ij} = x_{ij} + \phi_{ij}(x_{ij} - x_{kj}) \quad (5)$$

$j \in (1, 2, \dots, D)$ and $k \in (1, 2, \dots, Ns)$

3) Greedy Selection: Greedy selection assesses the quality of new sources of food using nectar amounts. If it surpasses the current position, employed bees move. If fitness equals or exceeds $X_i$, the new source replaces it in the population.

4) Onlooker Bees: Onlooker bees choose a food source using employed bees' information. The selection probability ($p_i$) depends on the food source's fitness ($f_i$). They generate a new source following the employed bee method. The new source is evaluated, with a similar greedy selection process as employed bees.

$$p_i = 0.9 \frac{f_{min}}{f_i} + 0.1 \tag{6}$$

5) Scout Bees: Scout bees replace abandoned solutions after repeated unsuccessful trials. When a solution can't be improved, it becomes abandoned, and the employed bee becomes a scout. Scouts randomly generate new solutions using equation (7).

$$v_{ij} = x_{min\,j} + rand\,[0,1].\,(x_{max\,j} - x_{min\,j}) \tag{7}$$

For $j$ = 1, 2, …, D

6) The algorithm stops when it reaches a termination condition, reporting the best solution. Termination conditions include reaching the maximum number of iterations $I_{max}$ or lack of convergence. Divergence is declared if no penalty-free solution is found within 10% of the total iterations, indicating low convergence.

**Penalized Objective Function & Penalty Function**

To assess the suitability of a trial design and ascertain its closeness to the global optimum, a penalty function is employed to compute the eventual constraint violation. This penalty function encompasses various geometric constraints related to cross-sectional dimensions, structural deflection, internal forces, and seismic performance [33], [37]. Consequently, the penalty increases in direct proportion to constraint violations, and the optimal design entails minimal weight and zero penalties. The penalized objective function quantifies the quality of a solution and is expressed as:

$$\varphi(x) = F(X).\,[1 + KC]^{\varepsilon} \tag{8}$$

$\varphi(x)$ = Penalized Objective Function

K = Penalty function constant

ε = Penalty function exponent.

In this study, K=1.0 and ε=2

The constraint violation (Equation 9) is defined as:

$$C = \sum_{i=1}^{n} c_i \tag{9}$$

$c_i$ = represent a specific violation function.

This study imposes a total of 21 constraints to assess the structural frame's adequacy.

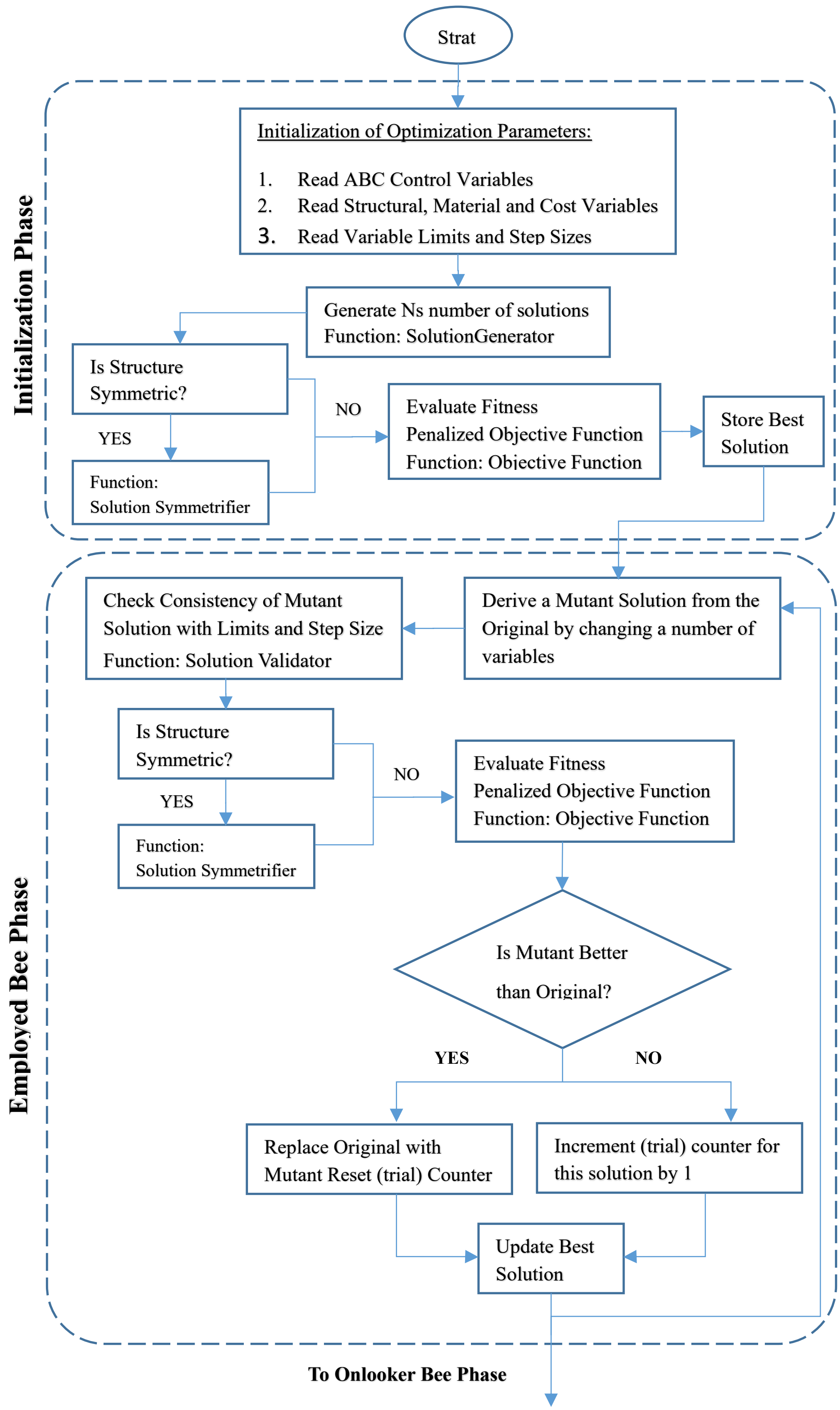
Strat
Initialization Phase
Initialization of Optimization Parameters:
1. Read ABC Control Variables
2. Read Structural, Material and Cost Variables
3. Read Variable Limits and Step Sizes
Generate Ns number of solutions
Function: SolutionGenerator
Is Structure Symmetric?
NO
YES
Evaluate Fitness
Penalized Objective Function
Function: Objective Function
Store Best Solution
Function:
Solution Symmetrifier
Employed Bee Phase
Check Consistency of Mutant Solution with Limits and Step Size
Function: Solution Validator
Derive a Mutant Solution from the Original by changing a number of variables
Is Structure Symmetric?
NO
YES
Evaluate Fitness
Penalized Objective Function
Function: Objective Function
Function:
Solution Symmetrifier
Is Mutant Better than Original?
YES
NO
Replace Original with Mutant Reset (trial) Counter
Increment (trial) counter for this solution by 1
Update Best Solution
To Onlooker Bee Phase

*Figure 1 Flowchart of the ABC Algorithm*

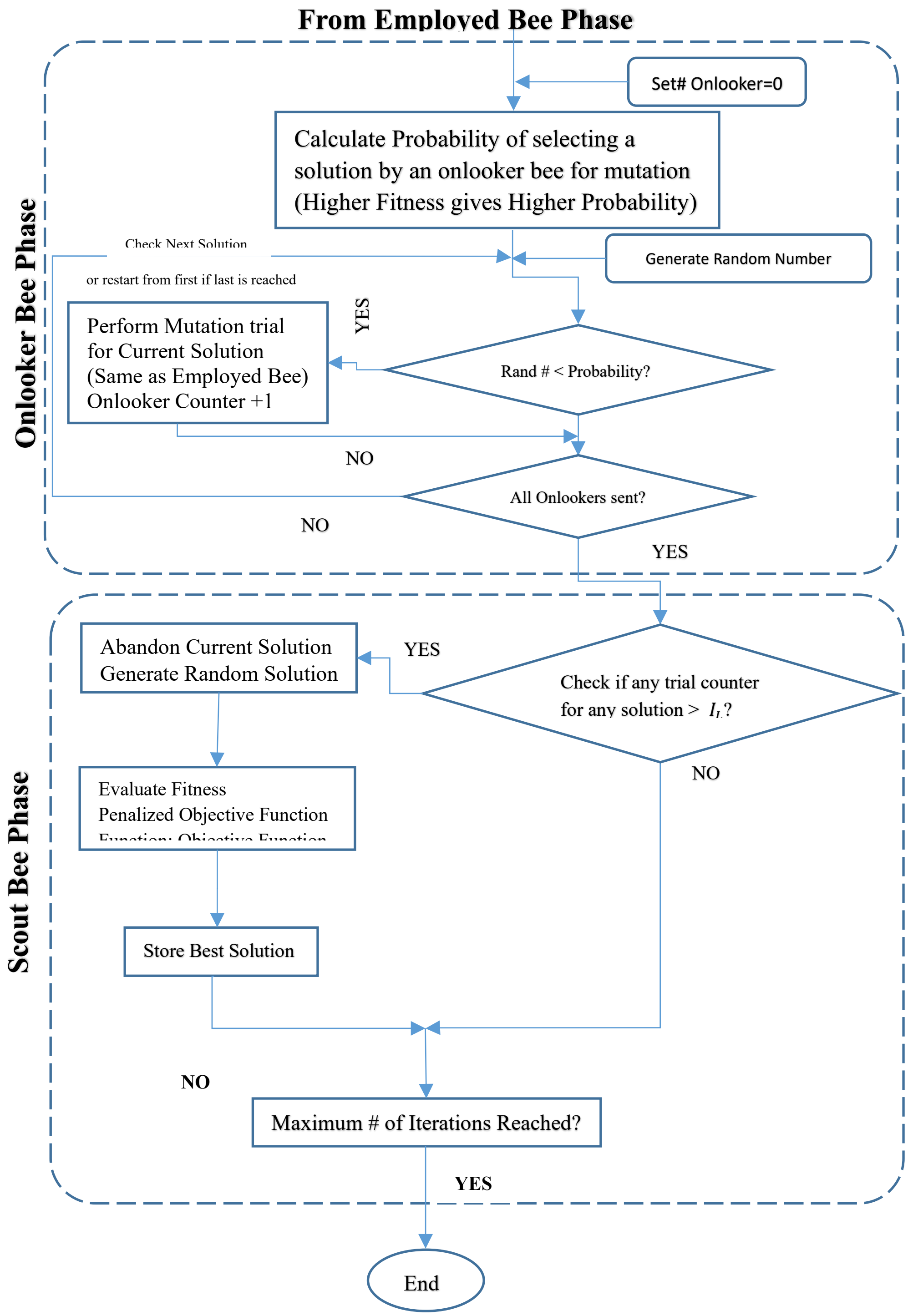

### 2.3. Design Parameters and Constraints for the Study

The primary objective of the optimization process is to minimize the weight of the structural frames. This objective is mathematically defined by the following equation [38]:

$$f(x) = \sum_{i=1}^{n}\left(\rho_{st_i} L_{st_i} A_{st_i}\right) + \sum_{j=1}^{m}\left(\rho_{cr_j} L_{cr_j} A_{cr_j}\right) \qquad (10)$$

'n' and 'm' = Total number of beams, columns, and shear walls.

$\rho_{st_i}$ = Steel density of the '$i$th' member,

$L_{st_i}$ = Length of '$i$th' member,

$A_{st_i}$ = Bar cross-section area of '$i$th' member.

j = Column number,

$\rho_{cr_j}$ = Concrete density,

$L_{cr_j}$ = Length of the '$jth$' column,

$A_{cr_j}$ = Concrete area of the '$j$th' column.

*Table 1 Allowable values at performance levels*

| Performance level | I-O | L-S | C-P |
|---|---|---|---|
| Maximum inter-story drift ratio % | 0.5 | 1 | 2 |

*Table 2 Allowable Rotation Values for Plastic Hinge in Concrete Columns.*

| $\frac{P}{A_g f_c'}$ | Transverse Reinforcement | $\frac{V}{b_w d\sqrt{f_c'}}$ | I-O | L-S | C-P |
|---|---|---|---|---|---|
| ≤ 0.1 | C | ≤ 3 | 0.005 | 0.015 | 0.02 |
| ≤ 0.1 | C | ≥ 6 | 0.005 | 0.012 | 0.016 |
| ≥ 0.4 | C | ≤ 3 | 0.003 | 0.012 | 0.015 |
| ≥ 0.4 | C | ≥ 6 | 0.003 | 0.01 | 0.012 |

*Table3 Allowable Rotation Values for Plastic Hinge in Concrete Beams.*

| $\frac{\rho - \rho'}{\rho_{bal}}$ | Transverse Reinforcement | $\frac{V}{b_w d\sqrt{f_c'}}$ | I-O | L-S | C-P |
|---|---|---|---|---|---|
| ≤ 0 | C | ≤ 3 | 0.01 | 0.02 | 0.01 |
| ≤ 0 | C | ≥ 6 | 0.005 | 0.01 | 0.02 |
| ≥ 0.5 | C | ≤ 3 | 0.005 | 0.01 | 0.02 |
| ≥ 0.5 | C | ≥ 6 | 0.005 | 0.005 | 0.015 |

*Table4 Allowable Rotation Values for Plastic Hinges in Reinforced Concrete Shear Walls.*

| $\frac{(A_S - A_S')f_y + P}{t_w l_w f_c'}$ | Border Confinement | $\frac{V}{t_w l_w\sqrt{f_c'}}$ | I-O | L-S | C-P |
|---|---|---|---|---|---|
| .1≤ 0 | YES | ≤ 3 | 0.005 | 0.01 | 0.015 |
| .1≤ 0 | YES | ≥ 6 | 0.004 | 0.008 | 0.01 |
| ≥ 0.25 | YES | ≤ 3 | 0.003 | 0.006 | 0.009 |
| ≥ 0.25 | YES | ≥ 6 | 0.0015 | 0.003 | 0.005 |
| .1≤ 0 | NO | ≤ 3 | 0.002 | 0.004 | 0.008 |
| .1≤ 0 | NO | ≥ 6 | 0.002 | 0.004 | 0.006 |
| ≥ 0.25 | NO | ≤ 3 | 0.001 | 0.002 | 0.003 |
| ≥ 0.25 | NO | ≥ 6 | 0.001 | 0.001 | 0.002 |

In this research, the OPENSEES platform [25] was employed to conduct pushover analysis as an integral component of the optimization process. During the performance-based design (PBD), certain constraints need to be evaluated within the optimization process. These constraints are related to the strength of the section and structural displacements. In the PBD approach, these constraints must be evaluated considering the effect of gravity loading, as described below [7], [8], [9]:

$$Q_G = 1.2Q_{DL} + 1.6Q_{LL} \tag{11}$$

Where $Q_{DL}$ and $Q_{LL}$ are dead and live loads, respectively.

**General Constraints of Frames:**

$$C_1 = \frac{\Delta_{rel} - \Delta_{rel.max}}{\Delta_{rel.max}} \geq 0, \tag{12}$$

$$\Delta_{rel,max} = \ Maximum\ story\ drift, \tag{13}$$

$$\Delta_{rel} = \ story\ drift \tag{14}$$

Column Constraints:

Axial Load Constraint: $C_2 = \frac{P_u - \phi P_n}{\phi P_n} \geq 0$ (15)

Bending Moment Constraint: $C_3 = \frac{M_u - \phi M_n}{\phi M_n} \geq 0$ (16)

Shear Force Constraint: $C_4 = \frac{V_u - \phi V_n}{\phi V_n} \geq 0$ (17)

Minimum Reinforcement Ratio Constraint: $C_5 = \frac{0.01 - \rho}{0.01} \geq 0$ (18)

Maximum Reinforcement Ratio Constraint: $C_6 = \frac{\rho - 0.08}{0.08} \geq 0$ (19)

Upper and Lower Column Dimension Constraint:

$$C_7 = \frac{b_{top} - b_{bottom}}{b_{bottom}} \geq 0 \quad (20)$$

$$C_8 = \frac{h_{top} - h_{bottom}}{h_{bottom}} \geq 0 \quad (21)$$

Beam-Column Dimension Constraint: $C_9 = \frac{b_{beam} - b_{column}}{b_{column}} \geq 0$ (22)

Stiffness and Slenderness Constraint of Columns:

$$C_{10} = \frac{b_{column} - h_{column}}{h_{column}} \geq 0 \quad (23)$$

$$C_{11} = \frac{\frac{kl_u}{r} - 100}{100} \geq 0 \quad (24)$$

Reinforcement Spacing Constraint: $C_{12} = \frac{S_{min} - S}{S_{min}} \geq 0$ (25)

Minimum Free Spacing between Longitudinal Reinforcements should be 10 mm, a minimum of 4 rebars should be placed in the four corners of the section, rebar arrangement should be symmetrical and on two opposite sides of the section, minimum concrete cover should be 10 mm, and the diameter of stirrups should be Φ10.

**Beam Constraints:**

Bending Moment Constraint: $C_{13} = \frac{M_u - \phi M_n}{\phi M_n} \geq 0$ (26)

Shear Force Constraint: $C_{14} = \frac{V_s - V_{s,max}}{V_{s,max}} \geq 0$ (27)

Area Constraint: $C_{15} = \frac{A_{st,min} - A_{st}}{A_{st,min}} \geq 0$ (28)

Deformability Factor Constraint: $C_{16} = \frac{0.004 - \varepsilon_t}{0.004} \geq 0$ (29)

Reinforcement Spacing Constraint: $C_{17} = \frac{S_{min} - S}{S_{min}} \geq 0$ (30)

Height Constraint: $C_{18} = \frac{h_{min}-h}{h_{min}} \geq 0$ (31)

To assess the seismic performance of the frame through pushover analysis when gravity and seismic loads are additive, it's necessary to obtain the gravity loads according to the following load combination (FEMA356 [8]):

$$Q_{PDB} = 1.1(Q_{DL} + Q_{LL}) \quad (32)$$

The permissible values for the constraints of plastic hinge rotations and inter-story relative displacements, according to FEMA356, are as follows.

$$C_{19} = \frac{d_j^i}{d_{all}^i} - 1 \leq 0 \qquad j = 1.2, \dots, ns \quad (33)$$

$$C_{20} = \frac{\theta_j^i}{\theta_{all}^i} - 1 \leq 0 \qquad j = 1.2, \dots, nc \quad (34)$$

$$C_{21} = \frac{\theta_k^i}{\theta_{all}^i} - 1 \leq 0 \qquad k = 1.2, \dots, nb \quad (35)$$

i = IO; LS; CP, and $\boldsymbol{d_{all}^i}$ = drift of the *j*th floor, the allowable drift value, ns denotes the number of floors, $\boldsymbol{\theta_j^i}$ and $\theta_k^i$ respectively denote the maximum plastic hinge rotation for columns and beams, $\boldsymbol{\theta_{all}^i}$ represents the allowable plastic hinge rotation for beams and columns, ns, nb and nc denote the number of shear walls, beams and columns [39], [41], [42] [7], [8], [9].

## 2.4. Methodology

In this study, we investigate the optimal design of 2D Reinforced Concrete Moment Frames and Shear Walls using a Performance-Based Design (PBD) approach, incorporating Pushover Analysis alongside the ABC (Artificial Bee Colony) Optimization Algorithm.

1. Structural Modeling: We comprehensively model the 2D Reinforced Concrete Moment Frames and Shear Walls and their associated parameters within the OpenSees software platform (Mazzoni et al. 2004). Our models strictly adhere to the seismic design guidelines outlined in FEMA 356 and ASCE 7.

2. Objective Function: The core objective of this study is to minimize the weight of the structural elements within the moment frames. In this context, we designate the cross-sectional profiles of these structural elements as the primary design variables of interest.

3. Optimization Framework: To conduct the optimization process, we leverage MATLAB [40] in conjunction with OpenSees. This integrated software environment empowers us to systematically search for the optimal design configuration of the 2D Reinforced Concrete Moment Frames.

4. Initial Population Generation: To commence the optimization procedure, we generate initial populations that conform to serviceability requirements. These populations adhere to both geometric constraints and FEMA performance-based seismic constraints, setting the foundation for the optimization process.

5. Artificial Bee Colony Algorithm: Within the ABC algorithm framework, we capitalize on the unique capabilities of this algorithm. It facilitates the exploration of the design space by employing artificial bees, allowing us to identify the most optimal design configuration that satisfies our predefined performance criteria.

6. Iterative Exploration of the Design Space: The core of the methodology revolves around iterative exploration. The ABC algorithm, finely tuned in the previous step, embarks on a systematic journey through the expansive design space. This exploratory phase generates a plethora of potential design solutions, all meticulously assessed for their adherence to seismic provisions.

7. Evaluation of Optimal Designs: The resultant optimal designs, stemming from the iterative exploration, undergo rigorous evaluation. This evaluation is facilitated through the application of performance indices meticulously designed to assess the seismic behavior of these solutions.

8. Iterative Refinement: Iteration is the linchpin of this methodology. The iterative process unfolds until satisfactory solutions emerge, solutions that seamlessly align with both the pre-established performance-based criteria and the constraints inherent in the seismic design problem.

This comprehensive and meticulously orchestrated methodology ensures that the seismic design optimization process is not only systematic but also highly effective. It takes into account the multifaceted nature of performance-based design while harnessing the power of the ABC algorithm to navigate the intricacies of seismic structural optimization.

## 3. Case Studies

### 3.1. Description of the 2D Reinforced Concrete Moment Frames

In this section, we embark on an insightful exploration of 2D Reinforced Concrete Moment Frames, featuring three illustrative examples: a 4-story, an 8-story, and a 12-story configuration. These moment frames have been meticulously designed to excel in seismic performance, making them pivotal subjects for our study. The visual representation of these frames, as showcased in Figure 3, offers a clear insight into their structural composition. To provide a holistic understanding, we begin by detailing the applied loads: a dead load of 600 kg/m² and an imposed live load of 200 kg/m² on the floor slabs. The structural elements of the frames, inclusive of columns and beams, are meticulously outlined in Figure 3. Notably, our primary reinforcement material comprises longitudinal steel bars characterized by a yield stress of 400 N/mm². The concrete employed within these frames is assumed to possess a compressive strength of 30 MPa. Our analytical journey commences with an equivalent static analysis, meticulously executed through the powerful OpenSees software. This analysis serves as the foundation for discerning structural demands. Within this process, we take into account various critical factors that influence seismic performance, including the P-Delta effect of columns and shear walls, the arrangement of reinforcements, and the strategic placement of cutoff points for reinforcements within the beams. For a precise representation of reinforcement arrangements, we employ the concept of fiber sections, expertly defined through non-linear beam-column elements imbued with elastic properties. These sections are ingeniously programmed to adapt dynamically, ensuring an accurate reflection of the specific requirements of each section. Additionally, cutoff points for reinforcements are ingeniously introduced, featuring the definition of four nodes for every beam, ultimately leading to the formulation of three finite elements. To accomplish the intricate computational aspects of our case studies, the integration of MATLAB (2021b) software into our methodology proves instrumental. This seamless synergy between OpenSees and MATLAB facilitates the optimization process and empowers us to delve deeply into the multifaceted realm of design configurations.
This combined description offers a holistic and in-depth understanding of the moment frames. It presents both the broader context and the technical intricacies, catering to a diverse readership with varying levels of expertise.

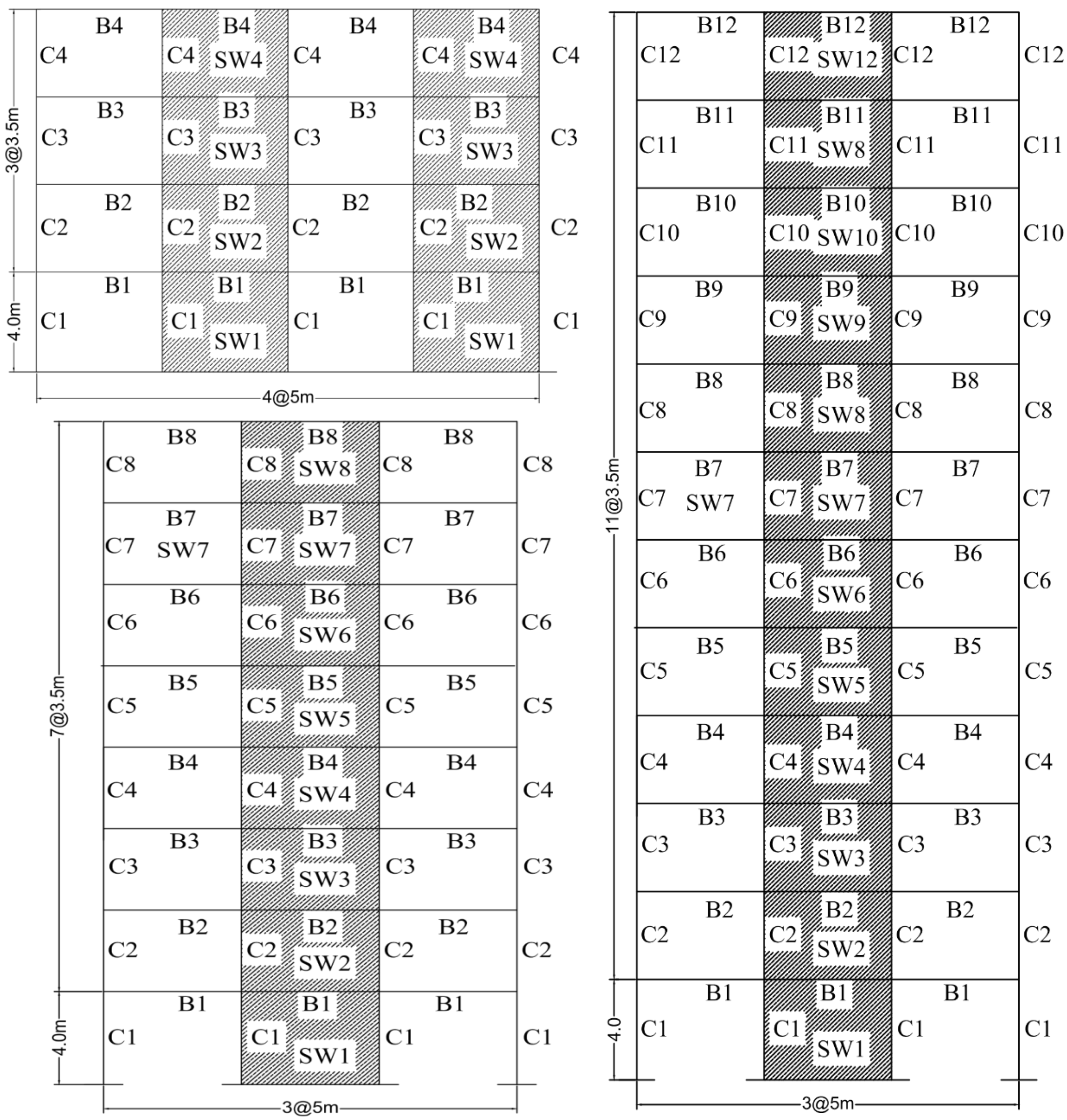

*Figure 2 Classification of Members in 4, 8, and 12-Story Reinforced Concrete Frames.*

### 3.2. Application of Performance-Based Design, Pushover Analysis, and ABC Optimization to the Case Studies

Our research journey continues as we delve into the practical application of Performance-Based Design (PBD), Pushover Analysis, and ABC (Artificial Bee Colony) Optimization within the context of this case study. To initiate this process, we first perform equivalent static analysis to determine structural demands, as mentioned earlier. The P-Delta effect of columns and shear walls, as well as reinforcement arrangements, are considered. For the latter, fiber sections are employed within the framework of non-linear beam-column elements. These sections adapt dynamically to different scenarios. Furthermore, reinforcement cutoff points in beams are incorporated by defining four nodes for each beam, resulting in three finite elements. The optimization process and all associated computational tasks are executed through MATLAB (2021b) software, thereby facilitating seamless communication and integration with OpenSees.

### 3.3. Performance Indices for Seismic Assessment

In this phase, we turn our attention to performance indices vital for seismic assessment. We begin by establishing the necessary structural parameters and loading conditions. Dead loads and live loads are accounted for, with values of 600 kg/m² and 200 kg/m², respectively. The axial loads of the shear wall within the dual system are considered concentrated loads equivalent to distributed loads. Material properties encompass concrete properties and steel properties

Load combinations are constructed based on the ACI 318, resulting in various cases denoted as:

$$U = 1.2D + 1.6L \tag{36}$$
$$U = 1.2D + 1.0L \pm 1.4E \tag{37}$$
$$U = 0.9D \pm 1.4E \tag{38}$$

These cases encompass different combinations of dead load (D), live load (L), and earthquake load (E). The allowable drift ratio is established at 0.0045, following the ASCE 7-10 code.

Our optimization approach employs the ABC (Artificial Bee Colony Algorithm) for optimizing a 4-story, an 8-story and a 12-story 2D RC dual system. This system consists of column groups, beam groups, and shear walls. Consequently, we have several design variables representing the unit weight of members in our optimization process.

Optimized structural member values derived from this process are compiled in Table 8, where it's important to note that constraints are violated by a negligible margin. This violation pertains to the beam stress constraint. Furthermore, Tables 5-7 provide the constraints' values, encompassing aspects like the convergence history of the optimization process for frames, and optimized concrete elements for all frames. Drift ratios for each story of all frames, specifically for the critical load combination, are visually depicted in Figures 9,12 and 13. This comprehensive analysis helps us evaluate the structural performance under seismic conditions and provides insights into the optimization process's effectiveness in achieving our defined objectives.

## 4. Results and Discussion

### 4.1. Comparative Analysis of Different Seismic Design Configurations:

This investigation explores three distinct examples, featuring 4-story RC 4-bays with shear walls on the 2nd and 4th bays, 8-story RC 3-bays with shear walls in the 2nd bay, and 12-story RC 3-bays with shear wall in the 2nd bay. These frames, illustrated in Figure 3, have been meticulously designed for optimal seismic performance. The floor slabs bear a dead load of 600 kg/m² and an imposed live load of 200 kg/m².

Longitudinal steel bars with a yield stress of 400 N/mm² serve as the primary reinforcement material. The concrete used is assumed to have a compressive strength of 30 MPa.

*Table5 Beam Cross-section Database.*

| Beam Number | Depth (mm) | Width (mm) | Number and Diameter of Bottom Reinforcement Bars | Number and Diameter of Top Reinforcement Bars |
|---|---|---|---|---|
| 1 | 300 | 300 | 3Φ16 mm | 3Φ16 mm |
| 2 | 300 | 300 | 3Φ18 mm | 3Φ18 mm |
| 3 | 300 | 300 | 4Φ20 mm | 4Φ20 mm |
| ... | ... | ... | ... | ... |
| 29 | 550 | 400 | 4Φ22 mm | 4Φ22 mm |
| 30 | 550 | 400 | 6Φ22 mm | 6Φ22 mm |
| 31 | 550 | 400 | 5Φ22 mm | 5Φ22 mm |

*Table6 Column Cross-section Database.*

| Column number | Column side length with square section (mm) | Number and diameter of bars |
|---|---|---|
| 1 | 300 | 8Φ16 mm |
| 2 | 300 | 8Φ18 mm |
| --- | --- | --- |
| 64 | 750 | 12Φ32 mm |
| 65 | 750 | 16Φ32 mm |

*Table7 Shear Wall Cross-section Database.*

| Shear wall number | $t_w(mm)$ | $t_f(mm)$ | $s_{sh}(mm)$ | $b_f(mm)$ | diameter of bars |
|---|---|---|---|---|---|
| 1 | 200 | 400 | 150 | 300 | Φ16 |
| 2 | 200 | 0 | 150 | 0 | Φ16 |
| --- | --- | --- | --- | --- | --- |
| 26 | 350 | 550 | 300 | 450 | Φ24 |
| 25 | 350 | 550 | 300 | 450 | Φ22 |

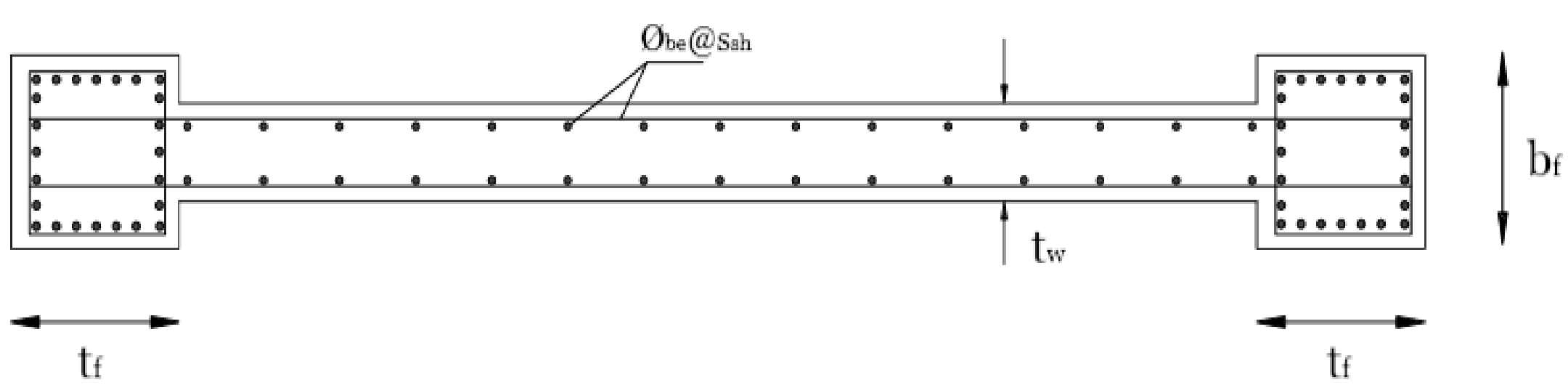

*Figure3 Concrete Shear Wall Components.*

In order to analyze the frames in OpenSees, fiber sections are utilized for modeling the sections. Beam objects are constructed using force-based nonlinear beam-column elements that incorporate two plastic hinges. The inclusion of second-order P-Delta effects is achieved through the implementation of the P-Delta coordinates transformation.

During the pushover analysis, lateral loads are applied to the frames based on Equation (7). This analysis allows for the evaluation of the structural response under gradually increasing loads.

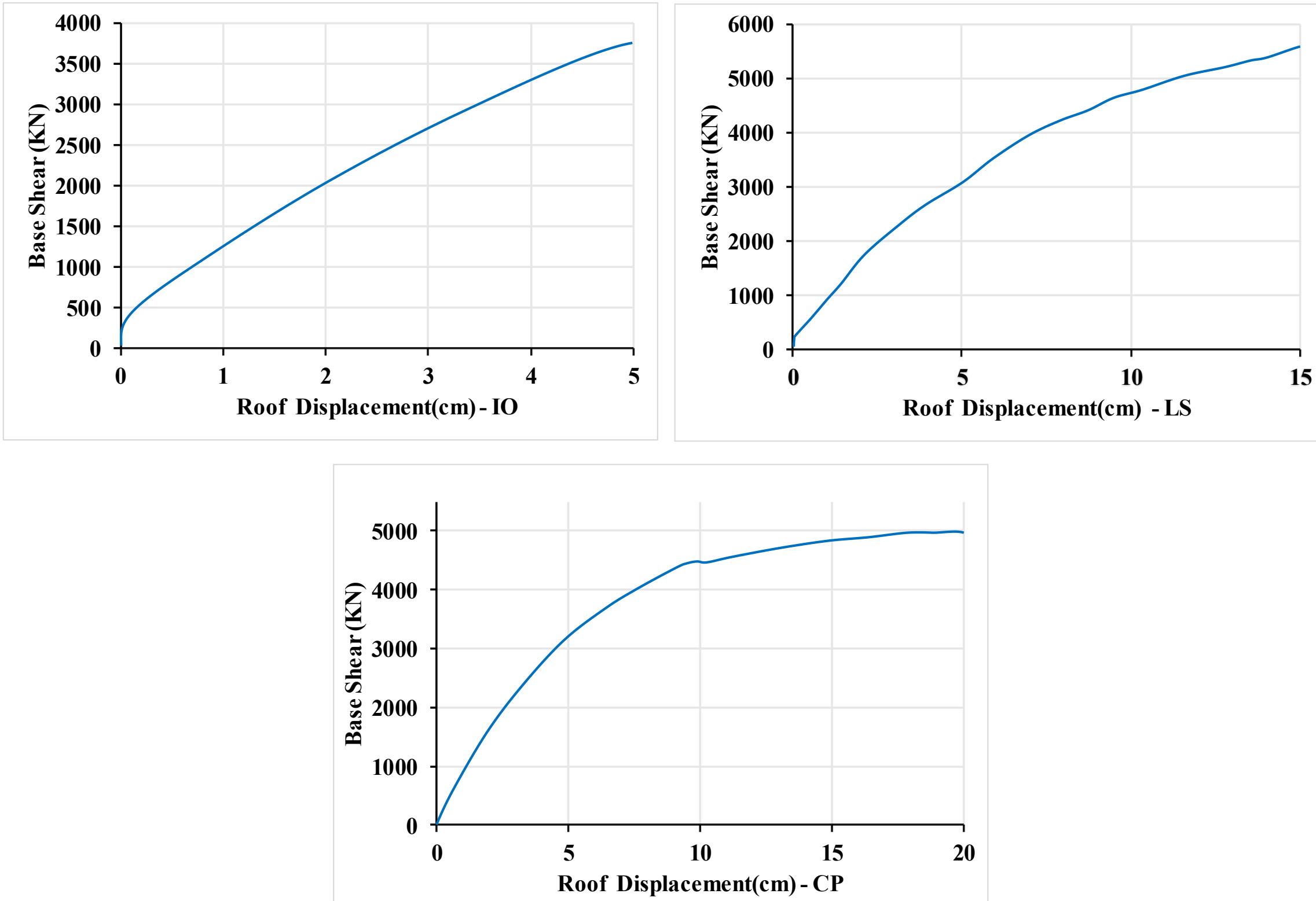

*Figure 4 Pushover Curve of a 4-story Reinforced Concrete Frame at Performance Levels IO, LS, CP.*

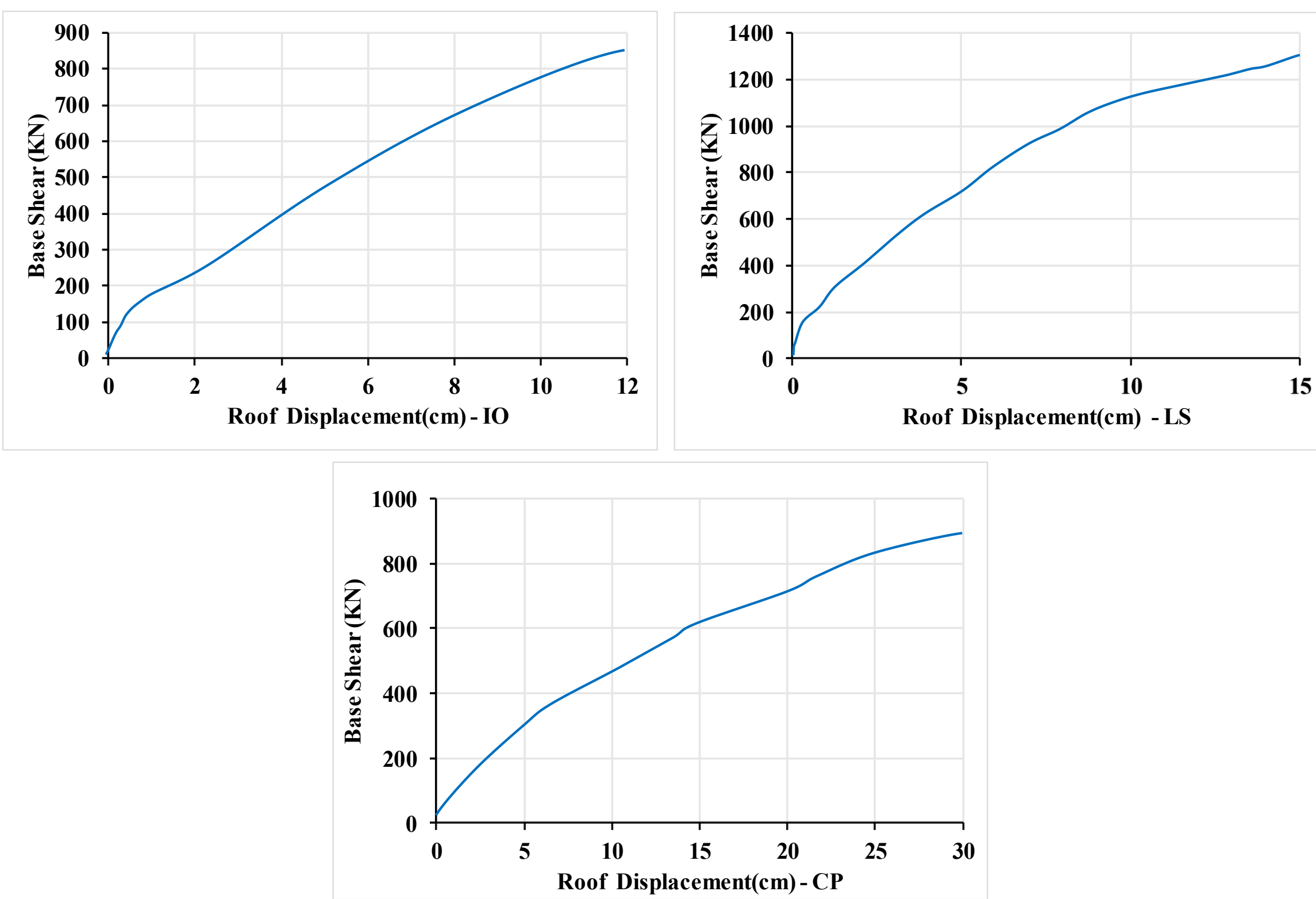

*Figure 5 Pushover Curve of an 8-story Reinforced Concrete Frame at Performance Levels IO, LS, CP.*

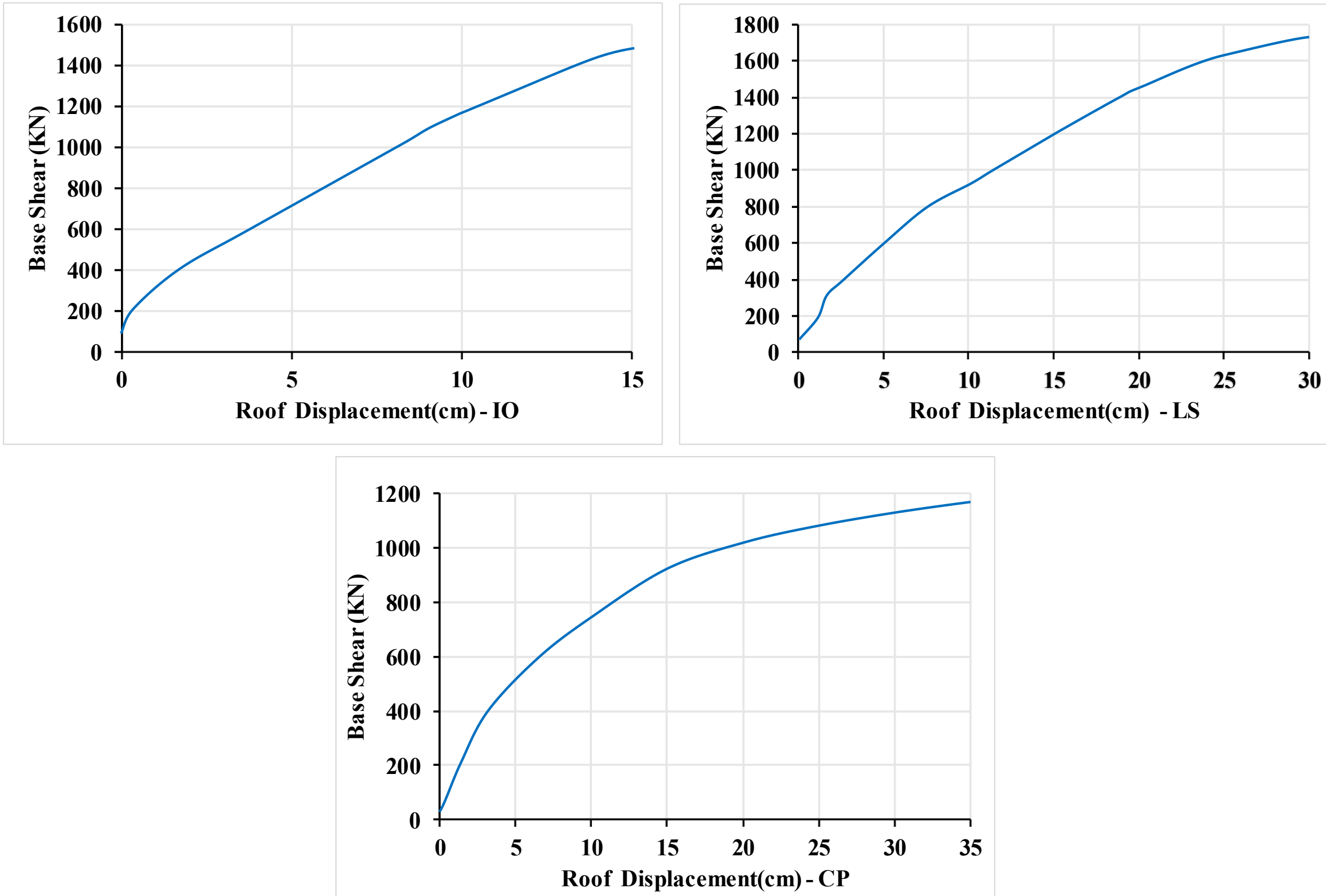

*Figure 6 Pushover Curve of a 12-story Reinforced Concrete Frame at Performance Levels IO, LS, CP.*

### 4.2. Identification of the Optimal Design using ABC Optimization:

The optimal design was determined using ABC (Artificial Bee Colony) optimization techniques. This method allowed us to explore a wide range of design configurations and parameters to identify the design that best met our performance criteria. After conducting numerous optimization runs, the following design parameters were found to yield the optimal results for each performance level:

*Table 8 Optimal Design Parameters for Different Performance Levels and Building Heights.*

| Optimal Design Parameters | 4-Story Building | 8-Story Building | 12-Story Building |
|---|---|---|---|
| Performance Level (IO) | | | |
| Roof Displacement (cm) | 5 | 5 | 15 |
| Shear Force Based On (KN) | 3850 | 3850 | 1485 |
| Number of Bees | 30 | 30 | 30 |
| Number of Repetitive Loads | 105 | 105 | 150 |
| Minimum Structure Weight (tons) | 182 | 182 | 321.12 |
| Allowable Relative Displacements (%) | 0.38% - 1.5% | 0.38% - 1.5% | 0.38% - 0.5% |
| Performance Level (LS) | | | |
| Roof Displacement (cm) | 15 | 25 | 30 |
| Shear Force Based On (KN) | 5600 | 1280 | 1720 |
| Number of Bees | 30 | 55 | 30 |
| Number of Repetitive Loads | 140 | 80 | 140 |
| Minimum Structure Weight (tons) | 4.178 | 5.212 | 293.09 |
| Allowable Relative Displacements (%) | 0.80% - 0.91% | 0.7% - 1% | 0.63% - 1% |
| Performance Level (CP) | | | |
| Roof Displacement (cm) | 20 | 30 | 35 |
| Shear Force Based On (KN) | 5000 | 930 | 1180 |
| Number of Bees | 30 | 30 | 30 |
| Number of Repetitive Loads | 140 | 140 | 150 |
| Minimum Structure Weight (tons) | 3.124 | 6.137 | 202 |
| Allowable Relative Displacements (%) | 1% - 1.75% | 1.2% - 1.8% | 1.42% - 2% |

### 4.3. Sensitivity Analysis of Design Parameters:

Roof Displacement Sensitivity: The sensitivity analysis revealed that roof displacement was highly sensitive to the chosen performance level. At the CP level, the roof displacement had to be significantly larger to prevent collapse, while at the IO level, a smaller displacement was acceptable due to the need for continuous use capability.

Shear Force Sensitivity: Shear force requirements varied significantly across performance levels. The IO level demanded the highest shear forces, reflecting the need for uninterrupted use, while the CP level required substantially lower shear forces due to the proximity to collapse.

Number of Bees Sensitivity: The number of bees used for analysis proved to be related to the number of repetitive loads. At the LS level, with more repetitive loads, a higher number of bees was required to obtain accurate results. Conversely, at the CP level, where fewer repetitive loads were present, a smaller number of bees sufficed for analysis.

These insights from the sensitivity analysis provide a valuable understanding of how specific design parameters should be tailored to meet particular performance goals in seismic design for each of the three buildings.

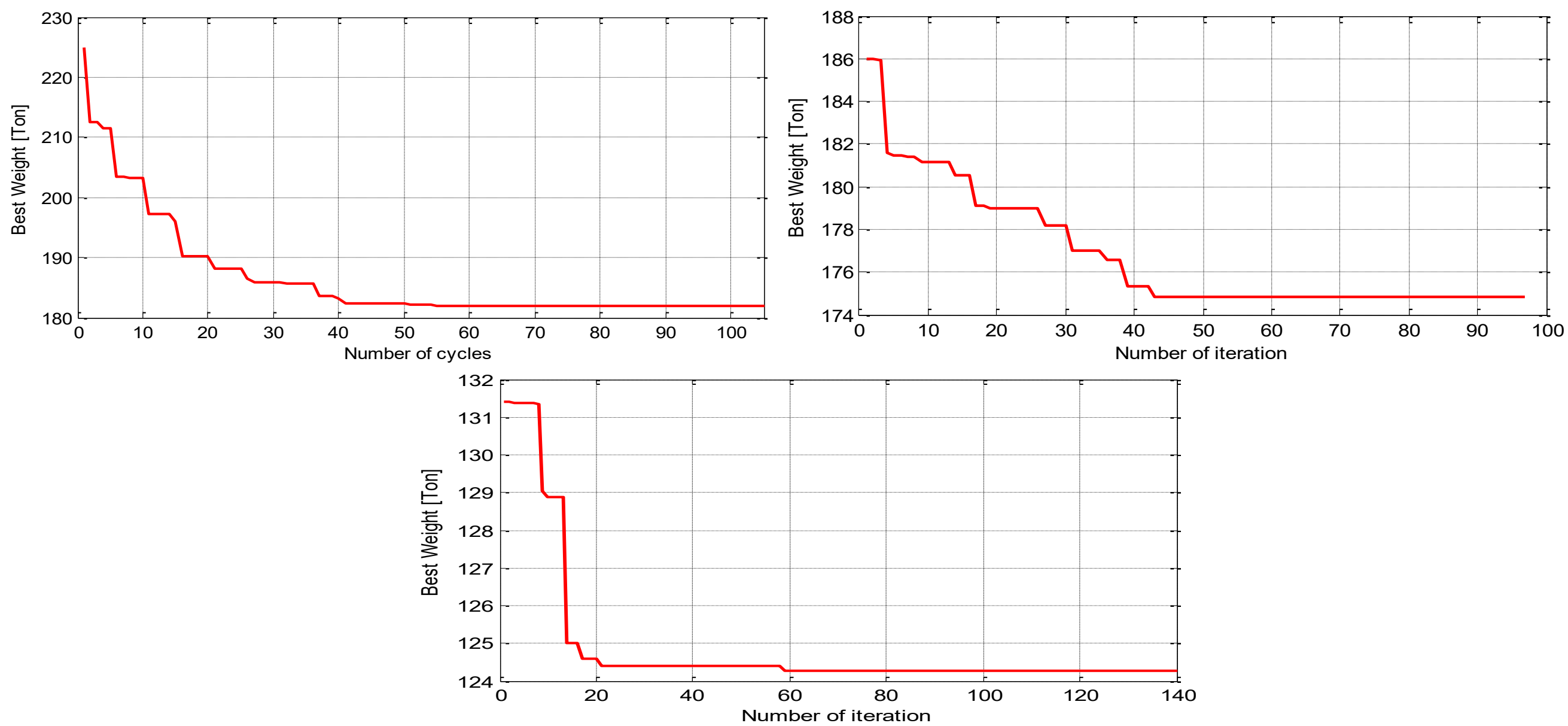

*Figure 7 Optimization Convergence History for a 4-Story Concrete Frame at Performance Levels IO, LS, CP.*

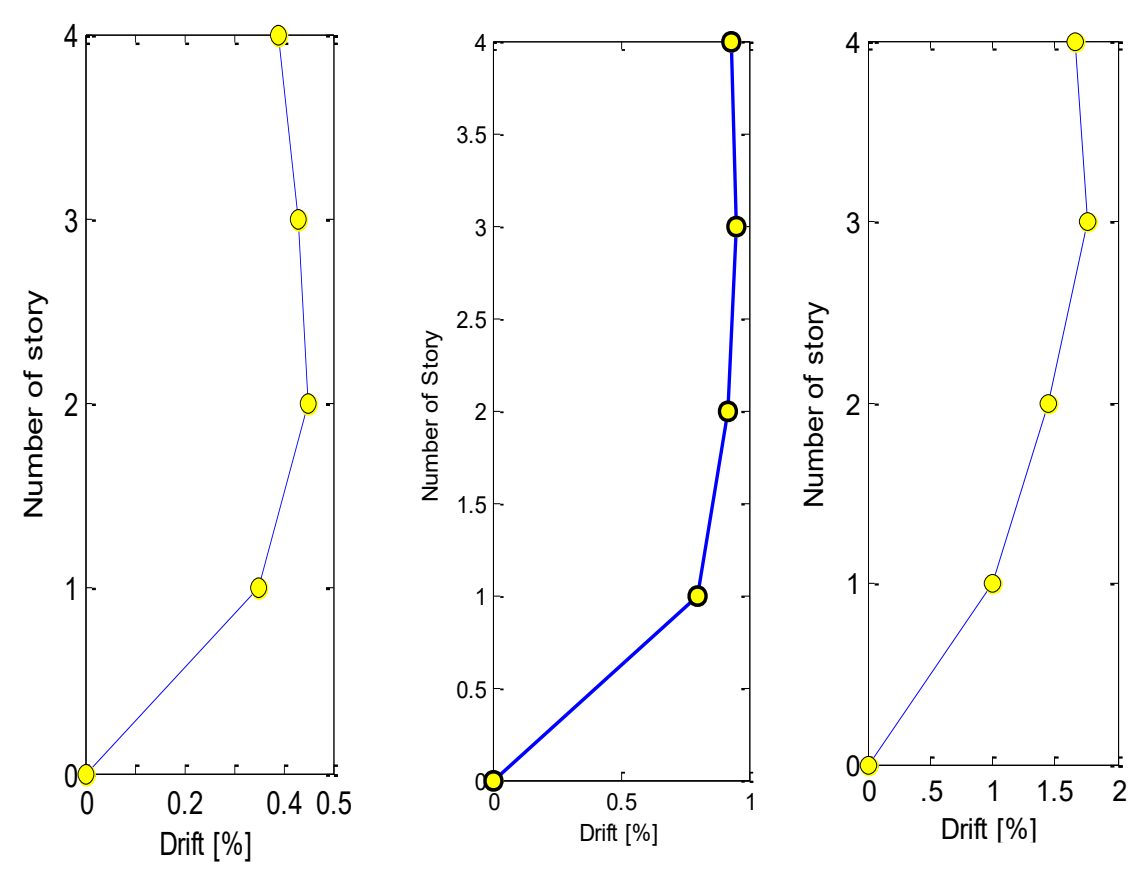

*Figure 8 Drift of 4-story Reinforced Concrete Structure, IO, LS, CP.*

*Table 9 Optimal Structural Sections for 4-story in IO, LS, CP.*

| Optimal Beam Sections IO | Optimal Column Sections IO | Shear wall typology numbers IO | Story |
|---|---|---|---|
| 3 | 22 | 2 | **4** |
| 11 | 36 | 6 | **3** |
| 15 | 36 | 10 | **2** |
| 17 | 38 | 12 | **1** |

| Optimal Beam Sections LS | Optimal Column Sections LS | Shear wall typology numbers LS | Story |
|---|---|---|---|
| 1 | 20 | 2 | **4** |
| 11 | 28 | 4 | **3** |
| 11 | 28 | 8 | **2** |
| 11 | 36 | 10 | **1** |

| Optimal Beam Sections CP | Optimal Column Sections CP | Shear wall typology numbers CP | Story |
|---|---|---|---|
| 2 | 20 | 2 | **4** |
| 1 | 28 | 2 | **3** |
| 1 | 28 | 4 | **2** |
| 2 | 36 | 6 | **1** |

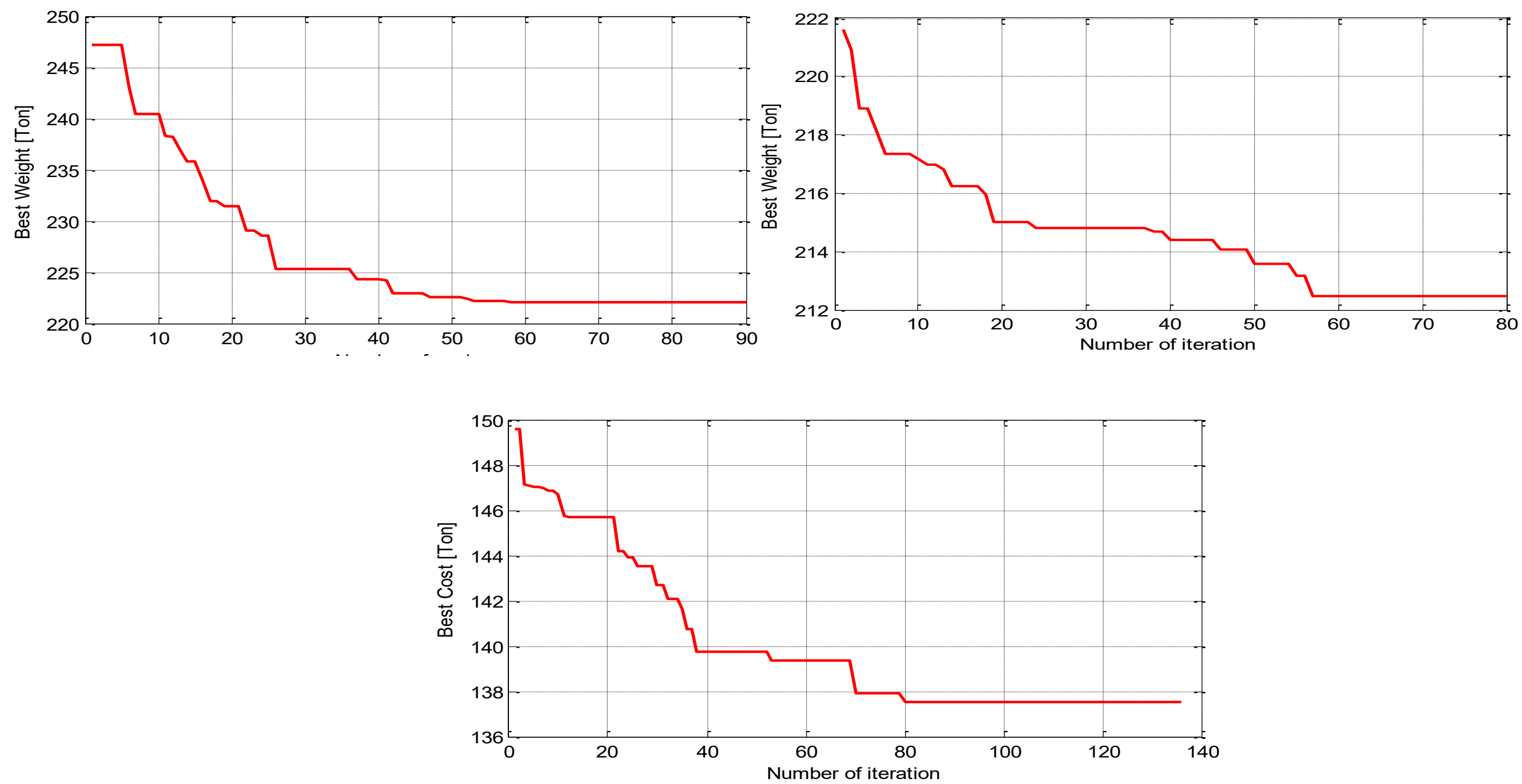

*Figure 9 Optimization Convergence History for an 8-Story Concrete Frame at Performance Levels IO, LS, CP.*

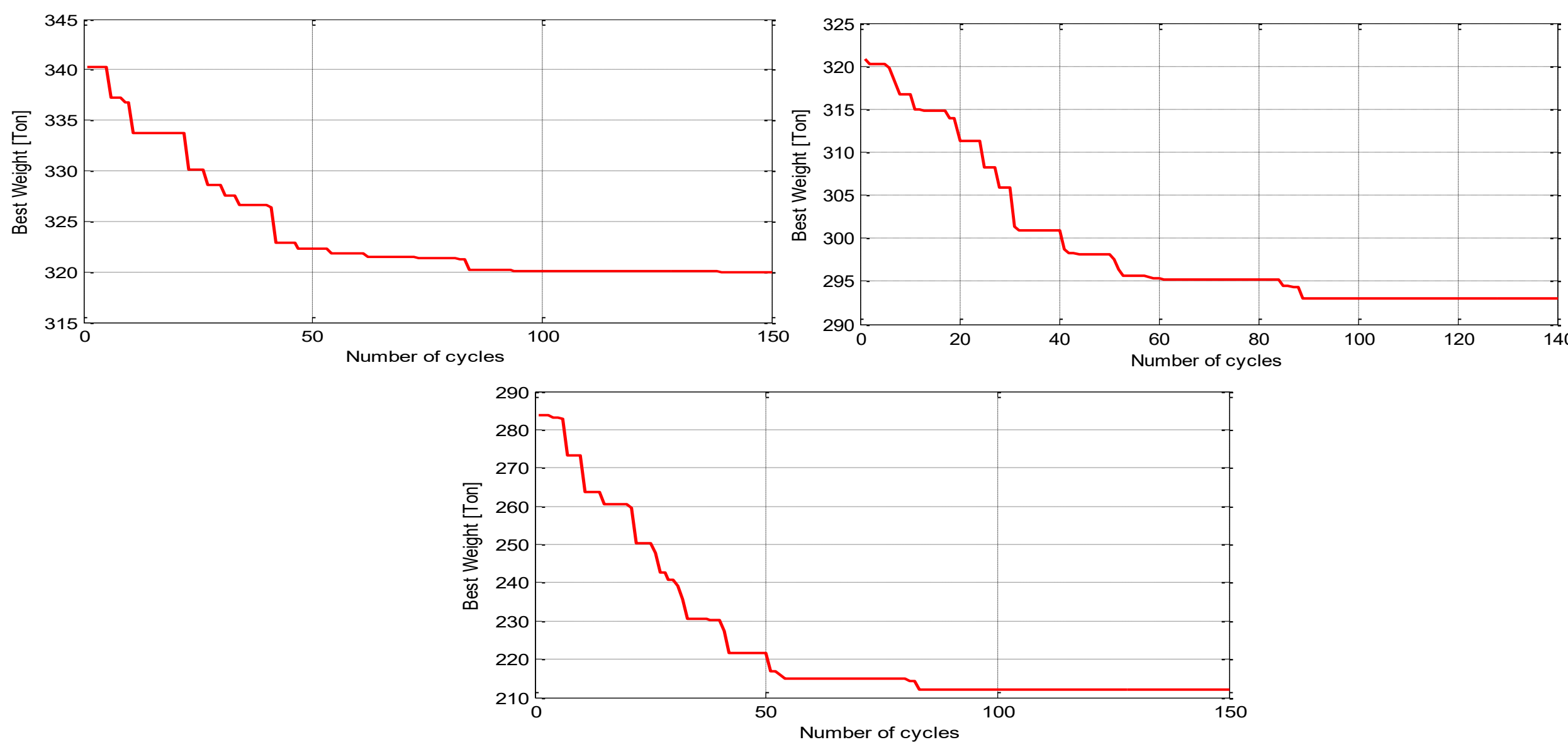

*Figure 10 Optimization Convergence History for a 12-Story Concrete Frame at Performance Levels IO, LS, CP.*

*Table 10 Optimal Structural Sections for 8-story in IO, LS, CP.*

| Optimal Beam Sections IO | Optimal Column Sections IO | Shear wall typology numbers IO | Story |
|---|---|---|---|
| 3 | 22 | 12 | **8** |
| 3 | 26 | 14 | **7** |
| 3 | 26 | 16 | **6** |
| 13 | 30 | 19 | **5** |
| 16 | 36 | 21 | **4** |
| 15 | 41 | 25 | **3** |
| 13 | 39 | 25 | **2** |
| 15 | 42 | 26 | **1** |

| Optimal Beam Sections LS | Optimal Column Sections LS | Shear wall typology numbers LS | Story |
|---|---|---|---|
| 1 | 20 | 10 | **8** |
| 1 | 20 | 14 | **7** |
| 3 | 23 | 14 | **6** |
| 12 | 26 | 16 | **5** |
| 15 | 28 | 18 | **4** |
| 13 | 36 | 19 | **3** |
| 13 | 36 | 22 | **2** |
| 15 | 38 | 22 | **1** |

| Optimal Beam Sections CP | Optimal Column Sections CP | Shear wall typology numbers CP | Story |
|---|---|---|---|
| 1 | 11 | 2 | **8** |
| 1 | 13 | 2 | **7** |
| 1 | 13 | 6 | **6** |
| 3 | 11 | 8 | **5** |
| 15 | 15 | 8 | **4** |
| 12 | 13 | 10 | **3** |
| 13 | 15 | 13 | **2** |
| 13 | 20 | 13 | **1** |

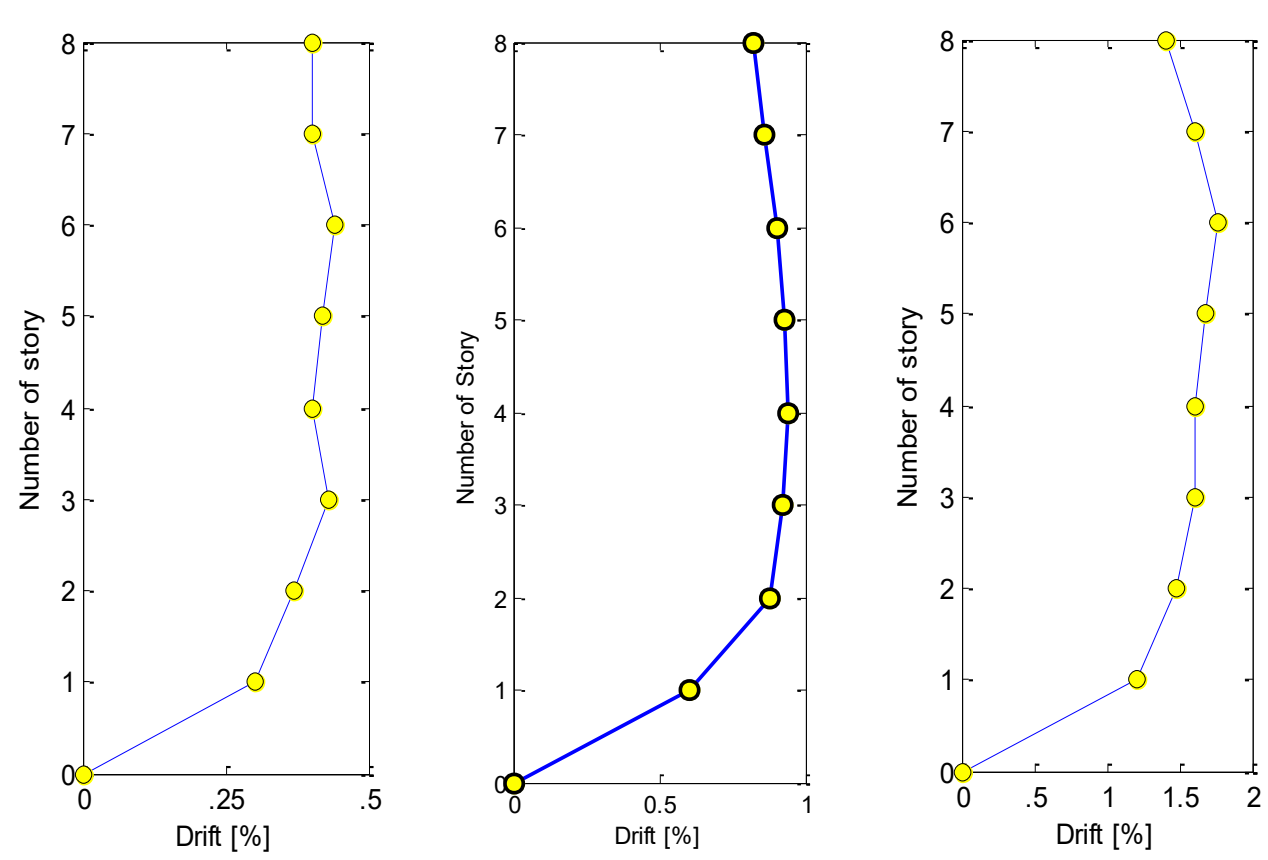

*Figure 11 Drift of the 8-story reinforced concrete structure at Performance Levels I-O, L-S, C-P.*

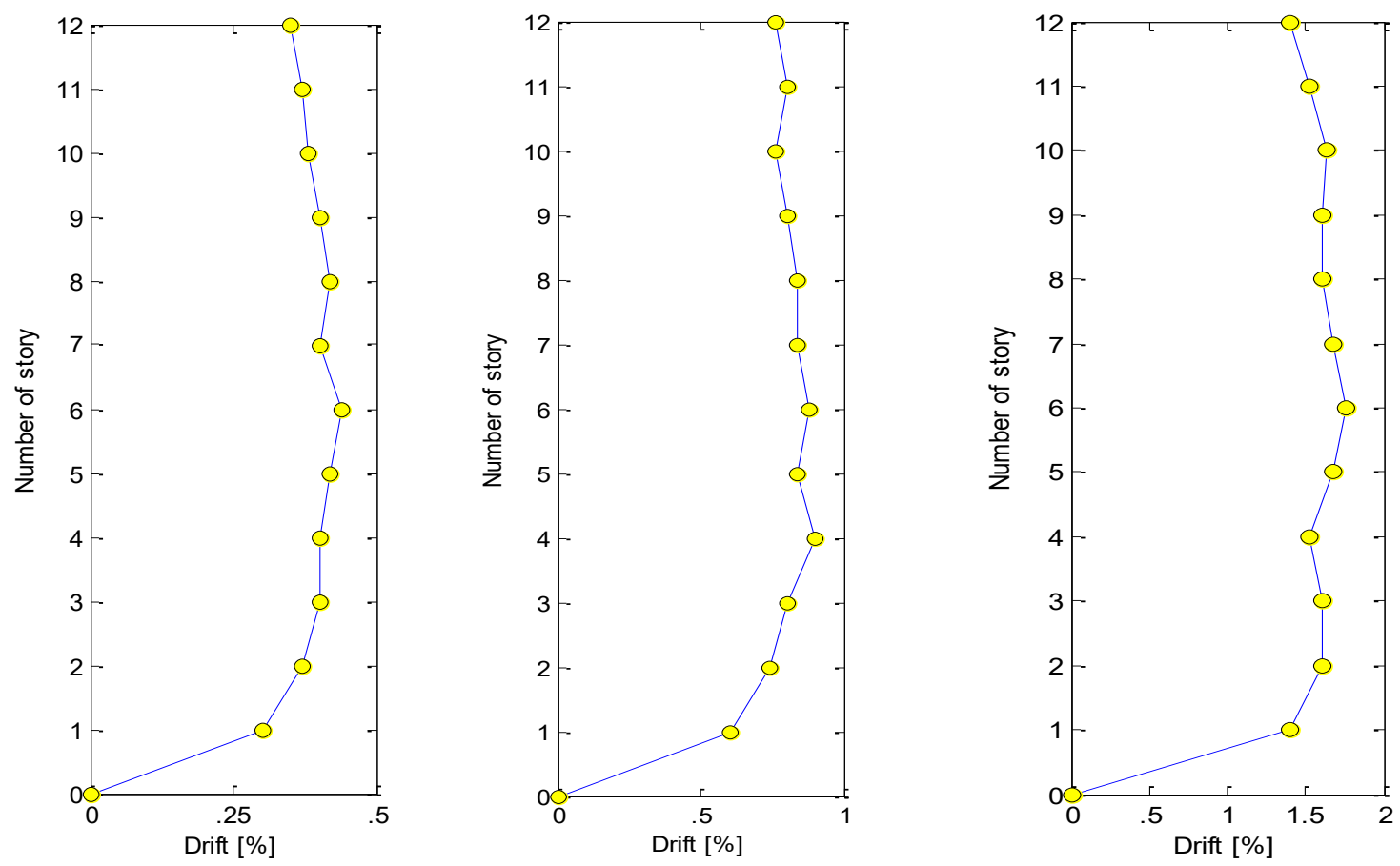

*Figure 12 Drift of the 12-story reinforced concrete structure at Performance Levels I-O, L-S, C-P.*

*Table 11 The optimized section numbers for the shear walls in the 12-story frame at Performance Levels I-O, L-S, C-P.*

| Optimal Beam Sections IO | Optimal Column Sections IO | Shear wall typology numbers IO | Story |
|---|---|---|---|
| 1 | 23 | 10 | **12** |
| 4 | 26 | 10 | **11** |
| 5 | 26 | 12 | **10** |
| 5 | 31 | 12 | **9** |
| 11 | 36 | 15 | **8** |
| 12 | 38 | 16 | **7** |
| 13 | 41 | 21 | **6** |
| 13 | 41 | 21 | **5** |
| 15 | 42 | 21 | **4** |
| 16 | 53 | 25 | **3** |
| 16 | 54 | 26 | **2** |
| 16 | 54 | 26 | **1** |

| Optimal Beam Sections LS | Optimal Column Sections LS | Shear wall typology numbers LS | Story |
|---|---|---|---|
| 1 | 21 | 10 | **12** |
| 1 | 26 | 10 | **11** |
| 4 | 26 | 12 | **10** |
| 2 | 30 | 13 | **9** |
| 5 | 34 | 15 | **8** |
| 5 | 38 | 18 | **7** |
| 11 | 41 | 18 | **6** |
| 12 | 41 | 19 | **5** |
| 13 | 42 | 19 | **4** |
| 13 | 44 | 20 | **3** |
| 15 | 48 | 21 | **2** |
| 16 | 48 | 22 | **1** |

| Optimal Beam Sections CP | Optimal Column Sections CP | Shear wall typology numbers CP | Story |
|---|---|---|---|
| 1 | 19 | 4 | **12** |
| 1 | 26 | 4 | **11** |
| 4 | 30 | 3 | **10** |
| 2 | 30 | 3 | **9** |
| 5 | 34 | 9 | **8** |
| 5 | 38 | 11 | **7** |
| 11 | 41 | 16 | **6** |
| 11 | 41 | 16 | **5** |
| 13 | 42 | 19 | **4** |
| 13 | 44 | 19 | **3** |
| 18 | 47 | 20 | **2** |
| 14 | 47 | 20 | **1** |

## 5. Conclusions

To complement the conclusions drawn from this extensive study, it's imperative to highlight the pivotal findings regarding the ABC Algorithm. This algorithm has unequivocally demonstrated its mettle as a rapidly converging and dependable tool for optimization. Notably, it consistently attains near-optimal solutions within a mere fraction of the total available iterations for a single run, signifying its remarkable efficiency. Moreover, its reliability and robustness shine through, regardless of the expansiveness of the design space. Throughout various test runs, the algorithm consistently maintains minimal standard deviations, bolstering its credibility as a steadfast optimizer.

In tandem with the algorithmic insights, the study underscores the immense potential inherent in shear wall-frame structures across diverse performance levels. By strategically optimizing these structures, engineers can harness the full capacity of frame members, thereby bolstering structural strength and ductility. Furthermore, the research reveals the critical nexus between a structure's displacement capacity up to its instability limit and its energy absorption and dissipation. As this displacement capacity increases, so does the structure's ability to dissipate energy effectively a crucial facet of seismic performance.

Transitioning from immediate occupancy to collapse prevention is an unequivocal catalyst for augmenting energy absorption and dissipation. This shift underscores the pivotal role of performance-based design in seismic engineering. Additionally, the study illuminates that as performance levels ascend from C-P to L-S and onward to I-O, structures become progressively robust and heavier, with heightened capacity. Optimized structures calibrated for immediate occupancy boast amplified stiffness and truncated periods. Notably, structural characteristics diverge concerning building height and configuration; for instance, a 4-story structure with two-span shear walls exhibits remarkable stiffness and capacity, while an 8-story counterpart featuring a single-span shear wall assumes a softer profile with a diminished capacity curve. Frames honed and optimized for collapse prevention exhibit pronounced nonlinear behavior, capitalizing on member rotation and deformation while judiciously deploying weaker sections to harness the nonlinear capacity of structural members. Collectively, these findings accentuate the promise of structural weight reduction across different performance levels, paving the way for substantial savings, especially in collapse prevention scenarios. The proximity of drift values to allowable thresholds further underscores the efficacy of the optimization methodology employed. These outcomes underscore the profound importance of performance-based design and sophisticated optimization techniques, like the ABC Algorithm, in elevating seismic performance to new heights.

In practice, the methodologies advanced through this research can be readily integrated into the seismic design of shear wall-frame structures, providing engineers with a pragmatic means to trim structural weight and attain optimal seismic design outcomes while aligning with industry guidelines. In summation, this study not only contributes to the ongoing advancement of seismic engineering but also arms professionals with practical tools and profound insights that not only enhance structural resilience but streamline design processes, ultimately paving the way for the creation of safer, more efficient structures in seismic-prone regions.

## 6. References

1. Das, T.K. and S. Choudhury, *Developments in the unified performance-based seismic design.* Journal of Building Pathology and Rehabilitation, 2023. **8**(1): p. 13.
2. Zakian, P. and A. Kaveh, *Topology optimization of shear wall structures under seismic loading.* Earthquake Engineering and Engineering Vibration, 2020. **19**: p. 105-116.
3. Wight, J.K., & Macgregor, J.G. , *Reinforced Concrete: Mechanics and Design*, ed. 6th. 2012, New Jersey: Pearson Education.
4. Banerjee, R., J. Srivastava, and N. Gupta, *Computational optimization of shear wall location in a C-shaped reinforced concrete framed building for enhanced seismic performance.* International Journal on Interactive Design and Manufacturing (IJIDeM), 2023: p. 1-10.
5. Mostofinejad, D., *Reinforced Concrete Structures*. Vol. Vol 2. 2021, Esfahan: Arkan Danesh.
6. SEAOC, *Vision 2000, Performance-based seismic engineering of buildings*. 2019, Structural Engineers Association of California: Sacramento (CA).
7. ATC, *Seismic evaluation and retrofit of concrete buildings*. 1996, Applied Technology Council: Redwood City, CA.
8. FEMA, *Prestandard and Commentary for the Seismic Rehabilitation of Buildings*, F. 356, Editor. 2000, Federal Emergency Management Agency: Washington, DC.
9. ASCE, *Seismic Evaluation and Retrofit of Existing Buildings*. 2022, American Society of Civil Engineers: Reston, VA.
10. Zou, X., C.M. Chan, G. Li, and Q. Wang, *Multiobjective optimization for performance-based design of reinforced concrete frames.* Journal of structural engineering, 2007. **133**(10): p. 1462-1474.
11. Sang-To, T., H. Le-Minh, S. Mirjalili, M.A. Wahab, and T. Cuong-Le, *A new movement strategy of grey wolf optimizer for optimization problems and structural damage identification.* Advances in Engineering Software, 2022. **173**: p. 103276.
12. Saka, M., *Optimum design of multistorey structures with shear walls.* Computers & structures, 1992. **44**(4): p. 925-936.
13. Ganzerli, S., C. Pantelides, and L. Reaveley, *Performance-based design using structural optimization.* Earthquake engineering & structural dynamics, 2000. **29**(11): p. 1677-1690.

14. Cheng, F.Y. and C. Pantelides, *Optimal placement of actuators for structural control*. 1988: National Center for Earthquake Engineering Research.
15. Fragiadakis, M. and N.D. Lagaros, *An overview to structural seismic design optimisation frameworks.* Computers & Structures, 2011. **89**(11-12): p. 1155-1165.
16. Grierson, D.E. and H. Moharrami, *Design optimization of reinforced concrete building frameworks*, in *Optimization of large structural systems*. 1993, Springer. p. 883-896.
17. Fadaee, M.J. and D.E. Grierson, *Design optimization of 3D reinforced concrete structures having shear walls.* Engineering with Computers, 1998. **14**: p. 139-145.
18. ACI, *Building Code Requirements for Structural Concrete (S.I.)*. 2019, American Concrete Institute.
19. Balling, R.J. and X. Yao, *Optimization of reinforced concrete frames.* Journal of structural engineering, 1997. **123**(2): p. 193-202.
20. Rajeev, S. and C. Krishnamoorthy, *Genetic algorithms-based methodologies for design optimization of trusses.* Journal of structural engineering, 1997. **123**(3): p. 350-358.
21. Sberna, A.P., F. Di Trapani, and G.C. Marano, *A new genetic algorithm framework based on Expected Annual Loss for optimizing seismic retrofitting in reinforced concrete frame structures.* Procedia Structural Integrity, 2023. **44**: p. 1712-1719.
22. Alavi, A., M. Dolatabadi, J. Mashhadi, and E. Noroozinejad Farsangi, *Simultaneous optimization approach for combined control–structural design versus the conventional sequential optimization method.* Structural and Multidisciplinary Optimization, 2021. **63**: p. 1367-1383.
23. Alemu, Y.L., B. Habte, T. Lahmer, and G. Urgessa, *Priority Criteria (PC) Based Particle Swarm Optimization of Reinforced Concrete Frames (PCPSO).* CivilEng, 2023. **4**(2): p. 679-701.
24. Mirrashid, M., H. Naderpour, D.-P.N. Kontoni, A. Jakubczyk-Gałczyńska, R. Jankowski, and T.N. Nguyen, *Optimized Computational Intelligence Model for Estimating the Flexural Behavior of Composite Shear Walls.* Buildings, 2023. **13**(9): p. 2358.
25. McKenna, F., Fenves, G. L., Scott, M. H., Jeremic, B., *OpenSees: Open system for earthquake engineering simulation*, 3.3.0, Editor., Pacific Earthquake Engineering Research Center, Berkeley University: California.
26. Jafari, A., M.R. Ghasemi, H. Akbarzadeh Bengar, and B. Hassani, *Seismic performance and damage incurred by monolithic concrete self-centering rocking walls under the effect of axial stress ratio.* Bulletin of Earthquake Engineering, 2018. **16**: p. 831-858.
27. Kaveh, A. and S.R. Ardebili. *Optimum design of 3D reinforced concrete frames using IPGO algorithm*. in *Structures*. 2023. Elsevier.
28. Kashani, A.R., A.H. Gandomi, K. Azizi, and C.V. Camp, *Multi-objective optimization of reinforced concrete cantilever retaining wall: a comparative study.* Structural and Multidisciplinary Optimization, 2022. **65**(9): p. 262.
29. Idels, O. and O. Lavan, *Performance based formal optimized seismic design of steel moment resisting frames.* Computers & Structures, 2020. **235**: p. 106269.
30. Hajirasouliha, I., P. Asadi, and K. Pilakoutas, *An efficient performance-based seismic design method for reinforced concrete frames.* Earthquake engineering & structural dynamics, 2012. **41**(4): p. 663-679.
31. Razavi, N. and S. Gholizadeh, *Seismic collapse safety analysis of performance-based optimally designed reinforced concrete frames considering life-cycle cost.* Journal of Building Engineering, 2021. **44**: p. 103430.
32. Gholizadeh, S. and V. Aligholizadeh, *Reliability-based optimum seismic design of RC frames by a metamodel and metaheuristics.* The Structural Design of Tall and Special Buildings, 2019. **28**(1): p. e1552.
33. Karaboga, D., *An idea based on honey bee swarm for numerical optimization*. 2005, Technical report-tr06, Erciyes university, engineering faculty, computer ….

34. Turan, S., İ. Aydoğdu, and E. Emsen, *Optimum Design of Elastic Continuous Foundations with The Artificial Bee Colony Method.* International Journal of Engineering and Applied Sciences, 2023. **15**(1): p. 36-51.
35. Yadhav, A., S. Gosavi, and M. Kulkarni, *Nonlinear behaviour of a reinforced concrete building subjected to blast load and optimisation using a meta-heuristic algorithm.* Asian Journal of Civil Engineering, 2023: p. 1-16.
36. Jahjouh, M., M. Arafa, and M. Alqedra, *Artificial Bee Colony (ABC) algorithm in the design optimization of RC continuous beams.* Structural and Multidisciplinary Optimization, 2013. **47**: p. 963-979.
37. Abdelmgeed, F., G. Ghallah, and A. Hamoda, *Optimum cost design of reinforced concrete beams using artificial bee colony algorithm.* Int J Adv Struct Geotech Eng, 2022. **6**: p. 110-129.
38. Moayyeri, N., S. Gharehbaghi, and V. Plevris, *Cost-based optimum design of reinforced concrete retaining walls considering different methods of bearing capacity computation.* Mathematics, 2019. **7**(12): p. 1232.
39. Shooli, A.R., A. Vosoughi, and M.R. Banan, *A mixed GA-PSO-based approach for performance-based design optimization of 2D reinforced concrete special moment-resisting frames.* Applied Soft Computing, 2019. **85**: p. 105843.
40. MathWorks, *MATLAB and Statistics Toolbox R2021b*. 2021, MathWorks: Natick, MA.
41. Faghirnejad, S., Kontoni, D.P.N., & Ghasemi, M.R. (2024). Performance-based optimization of 2D reinforced concrete wall-frames using pushover analysis and ABC optimization algorithm. Earthquakes and Structures, 27(4), 285–302. https://doi.org/10.12989/eas.2024.27.4.285
42. Faghirnejad, S., Kontoni, D.P.N., Camp, C.V., & Ghasemi, M.R. (2025). Seismic performance-based design optimization of 2D steel chevron-braced frames using ACO algorithm and nonlinear pushover analysis. Structural and Multidisciplinary Optimization, 68(1), 16. https://doi.org/10.1007/s00158-024-03948-y